\documentclass[english,aps,manuscript]{revtex4}
\usepackage[T1]{fontenc}
\usepackage[latin9]{inputenc}
\usepackage{textcomp}
\usepackage{relsize}
\usepackage{amsmath}
\usepackage{graphicx}
\usepackage{amssymb}

\makeatletter

\DeclareRobustCommand{\greektext}{%
  \fontencoding{LGR}\selectfont\def\encodingdefault{LGR}}
\DeclareRobustCommand{\textgreek}[1]{\leavevmode{\greektext #1}}
\DeclareFontEncoding{LGR}{}{}

\DeclareRobustCommand{\lyxmathsym}[1]{\ifmmode\begingroup\def\b@ld{bold}
  \def\rmorbf##1{\ifx\math@version\b@ld\textbf{##1}\else\textrm{##1}\fi}
  \mathchoice{\hbox{\rmorbf{#1}}}{\hbox{\rmorbf{#1}}}
  {\hbox{\smaller[2]\rmorbf{#1}}}{\hbox{\smaller[3]\rmorbf{#1}}}
  \endgroup\else#1\fi}

\providecommand{\tabularnewline}{\\}

\@ifundefined{definecolor}
 {\usepackage{color}}{}
\@ifundefined{definecolor}
 {\@ifundefined{definecolor}
 {\usepackage{color}}{}
}{}
\makeatother

\makeatother

\usepackage{babel}

\makeatother

\usepackage{babel}

\begin{document}

\title{A Search for the Fourth SM Family: Tevatron still has a Chance}

\author{M. SAHIN}

\email{m.sahin@etu.edu.tr}

\affiliation{TOBB University of Economics and Technology, Physics Division, Ankara,
Turkey}

\author{S. SULTANSOY}

\email{ssultansoy@etu.edu.tr}

\affiliation{TOBB University of Economics and Technology, Physics Division, Ankara,
Turkey}

\affiliation{Institute of Physics, National Academy of Sciences, Baku, Azerbaijan}

\author{S. TURKOZ}

\email{turkoz@science.ankara.edu.tr}

\affiliation{Ankara University, Department of Physics, Ankara, Turkey}
\begin{abstract}
Existence of the fourth family follows from the basics of the Standard
Model and the actual mass spectrum of the third family fermions. We
discuss possible manifestations of the fourth SM family at existing
and future colliders. The LHC and Tevatron potentials to discover
the fourth SM family have been compared. The scenario with dominance
of the anomalous decay modes of the fourth family quarks has been
considered in details. 
\end{abstract}
\maketitle

\section{INTRODUCTION}

Even though the Standard Model with three fermion families (SM3) accounts
for almost all of the large amount of the particle physics phenomena
\cite{PDG2008}, there are a number of fundamental problems which
cannot be addressed in the framework of the SM3: quark-lepton symmetry,
fermion's mass and mixing pattern, family replication and the number
of families, L-R symmetry breaking, electroweak scale etc. In addition,
SM3 contains unacceptably large number of arbitrary free parameters
put by hand: $19$ if the neutrinos are massless, $26$ if neutrinos
are Dirac particles and more than $30$ if neutrinos are Majorana
particles. Flavor Democracy Hypothesis (FDH), which is quite natural
in respect to the SM basics, provides a partial solutions to the above-mentioned
problems, namely: sheds light on fermion's mass and mixing pattern,
implies the number of SM families to be $4$ and reduces the number
of free parameters \cite{Fritzsch1,Datta,Celikel1995} (see also reviews
\cite{Sultansoy97-1,Sultansoy97-2,Sultansoy2000,Sultansoy2003,sultansoy2006,Sultansoy2009}
and references therein).

\emph{Historical analogy}: Let us emphasize the analogy of today's
SM fermions and parameters inflation with chemical elements inflation
in 19th century and hadron inflation in $1950-1960$. The both cases
have been clarified through four stages: systematics, predictions
confirmed, clarifying experiments, new basic physics level (see Table
I). We have added the last row to the Table in order to reflect present
situation in particle physics.

Let us remind that flavour physics met a lot of surprises. The first
example was discovery of $\mu$-meson (We were looking for $\pi$-meson
predicted by Yukawa but discovered the {}``heavy electron''). The
next example was represented by strange particles (later we understood
that they contains strange quarks). The story was followed by $\tau$-lepton,
c- and b-quarks discovered in 1970's. Actually, c-quark was foreseen
by GIM mechanism and quark-lepton symmetry and its mass was estimated
in the few GeV region, whereas the discovery of $\tau$-lepton and
b-quark was completely surprising for physicists. According to the
Standard Model they are the members of the third fermion family, which
was completed by the discovery of t-quark in 1995 at Tevatron. Actually,
we need at least three fermion families in order to handle CP-violation
within the SM \cite{Kobayashi}. CP violation is necessary for the
explanation of Barion Asymmetry of the Universe (BAU). Unfortunately,
SM with three fermion families does not provide actual magnitude for
BAU. Fortunately, the fourth SM family could provide additional factor
of order of $10^{10}$ and, therefore, solves the problem \cite{Hou}.

The status of the fourth SM family (SM4) was clearly emphasized at
dedicated international workshop held at CERN in September 2008. The
outcome of the workshop was published in a paper titled {}``Four
statements about the fourth generation\textquotedblright{} \cite{Holdom2008}.
These statements are:

1. The fourth generation is not excluded by EW precision data.

2. SM4 address some of the currently open questions.

3. SM4 can accommodate emerging possible hints of new physics.

4. LHC has the potential to discover or fully exclude SM4.

\vspace{6cm}

\begin{table}
\begin{tabular}{|c|c|c|c|c|}
\hline 
Inflation  & Systematic  & Confirmed  & Clarifying  & Fundamentals\tabularnewline
 &  & Predictions  & experiments  & \tabularnewline
\hline 
Chemical Elements  & Mendeleyev Periodic Table  & New elements  & Rutherford  & p, n, e\tabularnewline
\hline 
Hadrons  & Eight-fold Way  & New hadrons  & SLAC DIS  & quarks\tabularnewline
\hline 
SM fermions  & Flavor Democracy  & Fourth family ?  & LHC ?  & Preons ?\tabularnewline
\hline
\end{tabular}

\caption{Historical analogy}

\end{table}

\vspace{-6cm}

In our opinion the last statement is the most important one, because,
indirect manifestations could have many different explanations, the
existence of the fourth SM family will be proved with the direct discovery
of its quarks and leptons. Current experimental bounds on the masses
of the fourth SM family fermions are as follows \cite{PDG2008}:

$m_{u_{4}}>256$ GeV,

$m_{d_{4}}>128$ GeV (100\% CC decays); $m_{d_{4}}>199$ GeV (100\%
NC decays),

$m_{l_{4}}>100.8$ GeV,

$m_{\nu_{4}}>90.3$ GeV (Dirac type), $m_{\nu_{4}}(light)>80.5$ GeV
(Majorona type).

By this time almost all papers on the SM4 searches consider only SM
decay modes. However, it is possible that anomalous decay modes could
be dominant, if some criteria is met \cite{biz}. In this case, the
search strategy should be changed drastically and current low limits
from Tevatron experiments are not valid.

The scope of the paper is following: in Section 2 we give a brief
review of the Flavor Democracy Hypothesis and discuss possible manifestations
of the fourth SM family at existing and future colliders. Then, we
concentrate on the scenario with dominance of the anomalous decay
modes of the fourth family quarks. The criteria for this dominance
are considered in Section 3. In Section 4 we consider pair production
at the Tevatron and LHC with subsequent anomalous decays. Section
5 is devoted to investigation of anomalous resonant production of
the fourth family quarks with subsequent anomalous decays. Finally,
in Section 6 we present concluding remarks and recommendations.

\section{WHY THE FOUR SM FAMILIES ?}

First of all, the number of fermion families is not fixed by the SM.
But the asymptotic freedom restricts this number from above, namely,
$N\le8$. Then, the number of SM families with {}``massless'' neutrinos
(which means $m_{\lyxmathsym{\textgreek{n}}}<m_{Z}/2$) is determined
to be equal to $3$ by the LEP1 data. Therefore, number of families
could be any number between $3$ and $8$, inclusively. The most of
the free parameters (put by hand) in the SM comes from the Yukawa
interactions between the SM fermions and the Higgs doublet, which
provides fermion masses and mixings through Spontaneous Symmetry Breaking
(SSB). It should be noted that before the SSB, fermions with the same
quantum numbers are indistinguishable. Naturally, all Yukawa coupling
constants for indistinguishable fermions should be the same. This
is the first assumption of the Flavor Democracy Hypothesis. If there
is only one Higgs doublet all fundamental fermions (up and down type
quarks, charged leptons and neutrinos) should have the same Yukawa
coupling constants, since all fermions interact with the same Higgs
field. This is the second assumption of the FDH. After the SSB these
assumptions in the case of $N$ SM families result with $N-1$ fermion
families to be massless and the N'th family to be heavy and degenerate.
By taking into consideration masses of the third SM family, the FDH
implies at least the existence of the fourth SM family \cite{Fritzsch1,Datta,Celikel1995}.
In this case, the masses of the first three family fermions come from
the slight violation of the full democracy \cite{Datta2,Atag,Ciftci}.

There are two arguments against the existence of the fifth SM family
\cite{Sultansoy2000,sultansoy2006,Sultansoy2009}. The first one is
the large value of the t-quark mass: in the case of 5 SM families
the FDH gives $m_{t}<<m_{4}<<m_{5}$, but it contradicts to partial-wave
unitarity constraint $m_{Q}\le700$ GeV $\approx4$ $m_{t}$. The
second argument is the neutrino counting at the LEP1: data gives three
{}``massless\textquotedblright{} non-sterile neutrinos, whereas
in the case of the five SM families the FDH predicts this number to
be four.

The main reason why the HEP community has objected against the fourth
SM family so far comes from the incorrect interpretation of the electroweak
precision data. This interpretation since 1990's has been included
into PDG reports published bi-annually in leading HEP journals. It
should be noted that recent opinion \cite{Langacker2010} of the writers
of the corresponding part of PDG reports is not as strict as it was.
Actually in a number of papers published during the last decade \cite{okun2000,He2001,Gerard,Kribs2007,Lenz,chanov,Hashimoto,Lenz2,OPUCEM}
it has been shown that the precision data and the SM4 are not mutually
exclusive. It is interesting that the updated precision data is shifted
into the direction of SM4 predictions. For the investigation of the
compatibility of the precision data with the fourth SM family and
other physics beyond the SM3 a new code named OPUCEM \cite{OPUCEM,OPWEB}
has been developed very recently. Using this code we determined the
validity of SM4 with a given set of parameters, namely, $m_{u_{4}}=410$
GeV, $m_{d_{4}}=390$ GeV, $s_{34}=0.01$ (CKM mixing between fourth
and third SM family quarks), $m_{\lyxmathsym{\textgreek{n}}_{4}}(l)=105$
GeV for light Majorana neutrino, $m_{l_{4}}=450$ GeV, $m_{H}=290$
GeV and $m_{\lyxmathsym{\textgreek{n}}_{4}}(h)=2300$ GeV for heavy
Majorana neutrino. This set is favored by FDH if the common Yukawa
coupling for all SM4 fermions is equal to the $SU_{W}(2)$ gauge constant
$g_{W}$ ($m_{H}=290$ GeV corresponds to quartic coupling of Higgs
field equal to $g_{W}$). Result is $R=0.97$ which is two times better
than SM3 value $R=1.7$ (here $R=\lyxmathsym{\textgreek{D}}\lyxmathsym{\textgreek{q}}^{2}$
denotes the {}``distance'' from the central values of $S$ and $T$
parameters, for details see \cite{OPUCEM,OPWEB}).

Actually, there is an infinite number of SM4 points (analog of the
well known SUGRA points) which are in better agreemnet with precision
EW data than the SM3. In Table II we present three of them. In Figure
1 we present these points in S-T plane together with SM3 predictions.
It is seen that SM4 points are closer to central values of S and T
parameters.

\begin{table}
\begin{tabular}{|c|c|c|c|c|}
\hline 
SM4 points  & 1  & 2  & 3  & SM3\tabularnewline
\hline
\hline 
$m_{u_{4}}$, GeV  & $410$  & $440$  & $440$  & -\tabularnewline
\hline 
$m_{d_{4}}$, GeV  & $390$  & $390$  & $390$  & -\tabularnewline
\hline 
$m_{l_{4}}$, GeV  & $450$  & $390$  & $390$  & -\tabularnewline
\hline 
$m_{\nu_{4}}$(L),GeV  & $105$  & $91$  & $95$  & -\tabularnewline
\hline 
$m_{\nu_{4}}$(H), GeV  & $2300$  & $2900$  & $2900$  & -\tabularnewline
\hline 
$m_{H}$, GeV  & $290$  & $250$  & $115$  & $115$\tabularnewline
\hline 
$s_{34}$  & $0.01$  & $0.02$  & $0.02$  & -\tabularnewline
\hline 
$R$  & $0.97$  & $0.56$  & $0.036$  & $1.7$\tabularnewline
\hline 
$S$  & $0.17$  & $0.15$  & $0.09$  & $0$\tabularnewline
\hline 
$T$  & $0.19$  & $0.16$  & $0.12$  & $0$\tabularnewline
\hline
\end{tabular}

\caption{S, T and R parameters for there SM4 points and SM3.}

\end{table}

\begin{figure}
\includegraphics[scale=0.5]{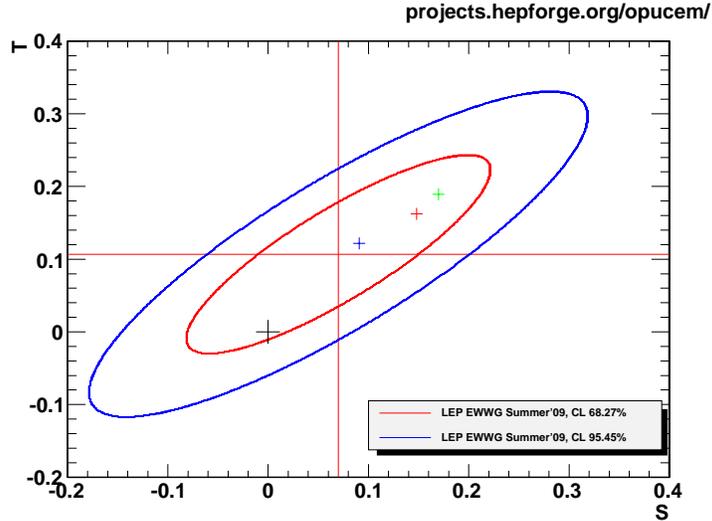}

\caption{SM3 and three SM4 points in S-T plane. The 1 and 2\textgreek{sv}
error ellipses represent the 2009 results of the U = 0 fi{}t from
LEP EWWG. Black crucifix corresponds to SM3 with $m_{H}=115$ GeV;
green, red and blue crucifixes correspond to SM4 points 1, 2 and 3
from Table 2, respectively. }

\end{figure}

\subsection{Indirect manifestations}

The existence of the fourth SM family could lead to a number of different
manifestations \cite{Holdom2008}, such as essential contribution
to the baryon asymmetry of the Universe (SM3 case does not provide
enough amount of CP violation), explanation for a $~2.5\lyxmathsym{\textgreek{sv}}$
deviation from SM3 predictions on B-meson decays observed by Tevatron
and B-factories etc. It should be noted that these are not a validation,
but just an indication of the fourth SM family, since there are a
lot of models (including SUSY) which potentially could lead to the
same manifestations. However, the essential enhancement (from $9$
times at $m_{H}\approx150$ GeV to $4$ times $m_{H}\approx500$ GeV)
of Higgs boson production via gluon-gluon fusion at hadron colliders
\cite{ATLASNote980,ATLASTDR1999-2,Ginzburg98,Cakir1,EArik,CakirGolden,EArik2,ArikBPL2007,N.Becerici}
could not be provided by other models. This enhancement could give
to Tevatron an opportunity to discover Higgs boson before the LHC
\cite{EArik2,N.Becerici}. Very recent combined results of CDF (with
$4.8$ $fb^{-1}$) and D0 (with $5.4$ $fb^{-1}$) searches for a
standard model Higgs boson in the process $gg\rightarrow H\rightarrow W^{+}W^{-}$
exclude $131$ GeV $<m_{H}<204$ GeV region in the SM4 case \cite{TeV4FH}.
This excluded region will be essentially enlarged with the accumulated
luminosity, or Higgs boson will be observed at Tevatron if it has
appropriate mass. Moreover, simultaneous discovery of both the Higgs
boson and the fourth family neutrino is probable at early stages of
LHC operation or at the Tevatron \cite{Unel,Cuhadar,Whiteson}.

\subsection{Direct manifestations}

Obviously, the discovery of the fourth SM family may be only provided
by their production and observation at high energy colliders. The
fourth family quarks will be copiously produced in pairs at the LHC
\cite{Arik98,ATLASTDR1999-2,Sultansoy97-1} when the designed center
of mass energy and luminosity values is achieved. However, Tevatron
still has a chance to observe $u_{4}$ before the LHC if $u_{4}$
mass is less than $425$ GeV (current low limit is $340$ GeV from
CDF with $4.6$ $fb^{-1}$). If the fourth family quarks mix dominantly
with first two families, $u_{4}$ and $d_{4}$ quarks will give the
same signature and observation limit will be extended to $450$ GeV.

\begin{table}
\begin{tabular}{|c|c|c|c|c|}
\hline 
Colliders  & Beams  & $\lyxmathsym{\textsurd}s$, TeV  & $L$, $10^{32}$ $cm^{-2}s^{-1}$  & $L_{int}(2012),fb^{-1}$\tabularnewline
\hline
\hline 
Tevatron  & $p\bar{p}$  & $1.96$  & $3.5$  & $12$\tabularnewline
\hline 
LHC 1  & $pp$  & $7$  & $0.01\rightarrow1$  & $1$\tabularnewline
\hline 
LHC 2  & $pp$  & $10$  & $10$  & \tabularnewline
\hline 
LHC 3  & $pp$  & $14$  & $100$  & \tabularnewline
\hline 
QCD-E 1  & $ep,\gamma p$  & $1.4$  & $30$ & \tabularnewline
\hline 
QCD-E 2  & $ep,\gamma p$  & $1.98$  & $10$ & \tabularnewline
\hline 
Linac-LHC EF  & $ep,\gamma p$  & $3.74$  & $3$ & \tabularnewline
\hline 
ILC 1  & $e^{-}e^{+},\gamma e,\gamma\gamma$  & $0.5$  & $100$ & \tabularnewline
\hline 
ILC 2  & $e^{-}e^{+},\gamma e,\gamma\gamma$  & $0.8$  & $100$ & \tabularnewline
\hline 
CLIC 1  & $e^{-}e^{+},\gamma e,\gamma\gamma$  & $0.5$  & $100$ & \tabularnewline
\hline 
CLIC 2  & $e^{-}e^{+},\gamma e,\gamma\gamma$  & $1$  & $100$ & \tabularnewline
\hline 
CLIC 3  & $e^{-}e^{+},\gamma e,\gamma\gamma$  & $3$  & $100$ & \tabularnewline
\hline 
Muon collider  & $\mu\mu$  & $4$  & $100$ & \tabularnewline
\hline
\end{tabular}

\caption{Parameters of existing and planned TeV scale colliders}

\end{table}

In Table III we present the center of mass energies and luminosity
values of exsisting and planned TeV scale colliders (see \cite{Sultansoy97-2,Sultansoy2003,Sultansoy EPS2003,Sultansoy PAC2005}
and refs therein). Observational possibilities for fourth SM family
fermions at these colliders are presented in Tables IV and V. The
direct production of the fourth SM family quarks and leptons at TeV
scale colliders, namely, Tevatron, LHC, QCD Explorer, Linac-LHC Energy
Frontier, ILC, CLIC and muon collider have been considered in a number
of papers \cite{Sher2000,Ciftci2001,Hakan2001,Rena2002,Arik2002,Orhan2002-1,Orhan2002-2,Hande2002,Orhan2003-1,Orhan2003-2,CLIC2004,Rena2005,Rena2006,Holdom2007-1,Holdom2007-2,Hung2007,Ciftci2008,Ozcan2008-1,Aguila2008,Cakir2008,Ozcan2008-2,Pree2008,Rena2008,Rena2009-1,Rena2009-2,Cakir2009-1,I.T.Cakir2009,Cakir2009-2,Rajaraman2010,Holdom2010,Ozcan2008-3,Ozcan2009}
(this list includes publications appeared during last decade, see
also fourth family web pages \cite{SM4 web-1,SM4 web-2}). In Tables
VI and VII it is given the classification of these papers according
to colliders and processes considered.

\begin{table}
\begin{tabular}{|c|c|c|c|c|c|}
\hline 
Colliders  & Tevatron  & LHC  & \multicolumn{3}{c|}{ILC/CLIC}\tabularnewline
\hline
\hline 
Beams  & $p\bar{p}$  & $pp$  & $e^{+}e^{-}$  & $\gamma e$  & $\gamma\gamma$\tabularnewline
\hline 
$q_{4}(P)$  & if KA  & VG  & if KA, VG  & -  & if KA \tabularnewline
\hline 
$\overline{u}_{4}d_{4}(AP)$ & ?  & ?  & -  & -  & -\tabularnewline
\hline 
$q_{4}(S)$  & {}``large'' $V_{4i}$  & {}``mid\textquotedblright{} $V_{4i}$  & -  & -  & -\tabularnewline
\hline 
$q_{4}(S,A)$  & {}``low\textquotedblright{} $\lyxmathsym{\textgreek{L}}$, Res  & {}``mid\textquotedblright{} $\lyxmathsym{\textgreek{L}}$, Res  & {}``low\textquotedblright{} $\lyxmathsym{\textgreek{L}}$  & -  & {}``low\textquotedblright{} $\lyxmathsym{\textgreek{L}}$\tabularnewline
\hline 
$l_{4}(P)$  & ?  & ?  & if KA, VG  & -  & if KA, G\tabularnewline
\hline 
$\lyxmathsym{\textgreek{n}}_{4}(P)$  & ?  & G  & if KA, VG  & -  & -\tabularnewline
\hline 
$l_{4}\nu_{4}(AP)$  & ?  & G  & -  & -  & -\tabularnewline
\hline 
$l_{4}(S,A)$  & ?  & ?  & {}``low\textquotedblright{} $\lyxmathsym{\textgreek{L}}$  & {}``mid\textquotedblright{} $\lyxmathsym{\textgreek{L}}$, Res  & {}``low\textquotedblright{} $\lyxmathsym{\textgreek{L}}$\tabularnewline
\hline 
$\nu_{4}(S,A)$  & ?  & ?  & {}``low\textquotedblright{} $\lyxmathsym{\textgreek{L}}$  & -  & -\tabularnewline
\hline 
Scalar Quarkonia  & -  & ?  & -  & -  & if KA, G\tabularnewline
\hline 
Vector Quarkonia  & -  & ?  & if KA, G  & -  & -\tabularnewline
\hline 
Hadrons  & -  & ?  & if KA, G  & -  & if KA\tabularnewline
\hline
\end{tabular}

\caption{Production of the fourth SM family fermions at existing and planned
high energy colliders. Abbreviations are: P (Pair production), AP
(Associate Pair production), S (Single production through CKM mixings),
A (production through Anomalous interactions), KA (Kinematically Allowed),
Res (Resonant production), G (Good), VG (Very Good). $V_{4i}$ denotes
corresponding CKM matrix elements, \textgreek{L} denotes scale
of anomalous interactions. }

\end{table}

\vspace{10cm}

\begin{table}
\begin{tabular}{|c|c|c|c|c|c|}
\hline 
colliders  & \multicolumn{4}{c|||}{Linac-LHC} & Muon Collider\tabularnewline
\hline 
 & \multicolumn{2}{c|}{QCD Explorer} & \multicolumn{2}{c|||}{Energy Frontier} & \tabularnewline
\hline 
Beams  & $ep$  & $\gamma p$  & $ep$  & $\gamma p$  & $\mu\mu$\tabularnewline
\hline
\hline 
$q_{4}(P)$  & if KA & if KA  & G  & VG  & VG\tabularnewline
\hline 
$q_{4}(AP)$  & -  & -  & -  & -  & -\tabularnewline
\hline 
$q_{4}(S)$  & {}``large'' $V_{4i}$  & -  & {}``mid'' $V_{4i}$  & -  & -\tabularnewline
\hline 
$q_{4}(S,A)$  & {}``low'' $\Lambda$  & {}``mid'' $\Lambda$, Res  & {}``mid'' $\Lambda$  & {}``mid'' $\Lambda$, Res  & {}``low'' $\Lambda$\tabularnewline
\hline 
$l_{4}(P)$  & -  & -  & -  & -  &  VG\tabularnewline
\hline 
$\nu_{4}(P)$  & -  & -  & -  & -  &  VG\tabularnewline
\hline 
$l_{4}\nu_{4}(AP)$  & -  & -  & -  & -  & -\tabularnewline
\hline 
$l_{4}(S,A)$  & {}``low'' $\Lambda$  & -  & {}``mid'' $\Lambda$  & -  & {}``low'' $\Lambda$\tabularnewline
\hline 
$\nu_{4}(S,A)$  & {}``low'' $\Lambda$  & -  & {}``mid'' $\Lambda$  & -  & {}``low'' $\Lambda$\tabularnewline
\hline 
Scalar Quarkonia  & -  & -  & -  & -  & -\tabularnewline
\hline 
Vector Quarkonia  & -  & -  & -  & -  & VG\tabularnewline
\hline 
Hadrons  & -  & -  & -  & -  & VG\tabularnewline
\hline
\end{tabular}

\caption{Notations as in Table IV.}

\end{table}

\vspace{-10cm}

\begin{table}
\begin{tabular}{|c|c|c|c|c|c|}
\hline 
Colliders  & Tevatron  & LHC  & \multicolumn{3}{c|}{ILC/CLIC}\tabularnewline
\hline
\hline 
Beams  & $p\overline{p}$  & $pp$  & $ee$  & $\gamma e$  & $\gamma\gamma$\tabularnewline
\hline 
$u_{4}$ (P), SM decays  & \cite{Hung2007,Ozcan2008-3} & \cite{ATLASTDR1999-2,Arik98,Holdom2007-1,Holdom2007-2,Ozcan2008-1,Aguila2008,Ozcan2008-2}  & \cite{Rena2002,CLIC2004}  &  & \cite{Rena2002,CLIC2004}\tabularnewline
\hline 
$u_{4}$(P), Anom decays  &  &  &  &  & \tabularnewline
\hline 
$d_{4}$(P), SM decays  & \cite{Ozcan2008-3} & \cite{ATLASTDR1999-2,Holdom2007-2,Ozcan2008-1,Aguila2008,Ozcan2008-2,Holdom2010} & \cite{Rena2002,CLIC2004} &  & \cite{Rena2002,CLIC2004}\tabularnewline
\hline 
$d_{4}$ (P), Anom decays  & \cite{biz,Sher2000,Hung2007} &  &  &  & \tabularnewline
\hline 
$q_{4}$(AP)  &  &  &  &  & \tabularnewline
\hline 
$q_{4}$(S)  &  & \cite{Cakir2008} &  &  & \tabularnewline
\hline 
$q_{4}$(S, A), SM decays  & \cite{Orhan2002-2,Orhan2003-1} & \cite{Rena2008,I.T.Cakir2009} & \cite{Hande2002} &  & \tabularnewline
\hline 
$q_{4}$(S, A), Anom decays  & \cite{Orhan2002-1,Orhan2002-2,Orhan2003-1,Orhan2003-2} & \cite{AIP2007_1,AIP2007_2,Rena2008} & \cite{Hande2002,AIP2007_1,AIP2007_2} &  & \tabularnewline
\hline 
$l_{4}$(P)  &  &  & \cite{Rena2002,CLIC2004} &  & \cite{Rena2002,CLIC2004}\tabularnewline
\hline 
$\nu_{4}$(P)  & \cite{Rajaraman2010} & \cite{Unel,Cuhadar,Rajaraman2010} & \cite{Rena2002,CLIC2004,Rena2005} &  & \tabularnewline
\hline 
$l_{4}\nu_{4}$(AP)  &  & \cite{Ozcan2009} &  &  & \tabularnewline
\hline 
$l_{4}$(S, A)  &  & \cite{Pree2008} &  & \cite{Rena2009-1} & \tabularnewline
\hline 
$\nu_{4}$(S, A)  &  &  &  &  & \tabularnewline
\hline 
Scalar Quarkonia  &  & \cite{ATLASTDR1999-2,Arik2002} &  &  & \cite{Rena2002,CLIC2004}\tabularnewline
\hline 
Vector Quarkonia  &  &  & \cite{Hakan2001,Rena2002,CLIC2004} &  & \tabularnewline
\hline 
Hadrons  &  &  & \cite{Hakan2001} &  & \tabularnewline
\hline
\end{tabular}

\caption{The papers considered production of the fourth SM family fermions
at existing and planned high energy colliders.}

\end{table}

\vspace{-40cm}

\begin{table}
\begin{tabular}{|c|c|c|c|c|c|}
\hline 
Colliders  & \multicolumn{4}{c|||}{Linac-LHC} & muon collider\tabularnewline
\cline{2-6} 
\multicolumn{1}{|c|}{} & \multicolumn{2}{c|}{QCD Explorer} & \multicolumn{2}{c|||}{Energy Frontier} & \tabularnewline
\hline 
Beams  & $ep$  & $\gamma p$  & $ep$  & $\gamma p$  & $\mu\mu$\tabularnewline
\hline 
$u_{4}$(P), SM decays  &  &  &  &  & \cite{Ciftci2001}\tabularnewline
\hline 
$u_{4}$(P), Anom decays  &  &  &  &  & \tabularnewline
\hline 
$d_{4}$(P), SM decays  &  &  &  &  & \cite{Ciftci2001}\tabularnewline
\hline 
$d_{4}$ (P), Anom decays  &  &  &  &  & \tabularnewline
\hline 
$q_{4}$(AP)  &  &  &  &  & \tabularnewline
\hline 
$q_{4}$(S)  & \cite{Cakir2009-1,Cakir2009-2} &  &  &  & \tabularnewline
\hline 
$q_{4}$(S, A), SM decays  &  &  &  &  & \tabularnewline
\hline 
$q_{4}$ (S, A), Anom decays  & \cite{AIP2007_1,AIP2007_2,Rena2009-2} &  &  &  & \tabularnewline
\hline 
$l_{4}$(P)  &  &  &  &  & \cite{Ciftci2001}\tabularnewline
\hline 
$\nu_{4}$(P)  &  &  &  &  & \cite{Ciftci2001}\tabularnewline
\hline 
$l_{4}\nu_{4}$ (AP)  &  &  &  &  & \tabularnewline
\hline 
$l_{4}$(S, A)  & \cite{Rena2006} &  & \cite{Rena2006} &  & \tabularnewline
\hline 
$\nu_{4}$ (S, A)  & \cite{Ciftci2008} &  & \cite{Ciftci2008} &  & \tabularnewline
\hline 
Scalar Quarkonia  &  &  &  &  & \tabularnewline
\hline 
Vector Quarkonia  &  &  &  &  & \cite{Ciftci2001}, \cite{Hakan2001}\tabularnewline
\hline 
Hadrons  &  &  &  &  & \cite{Hakan2001}\tabularnewline
\hline
\end{tabular}

\caption{The papers considered production of the fourth SM family fermions
at existing and planned high eergy colliders (cont.)}

\end{table}

\section{ANOMALOUS DECAY MODES}

The effective Lagrangian for anomalous magnetic type interactions
of the fourth family quarks is given as \cite{Cabibbo,Hagiwara,Orhan2003-1}:

\begin{equation}
L={\displaystyle \underset{{\scriptstyle q_{i}}}{{\displaystyle \sum}}\frac{\kappa_{\gamma}^{q_{i}}}{\Lambda}e_{q}g_{e}\bar{q_{4}}\sigma_{\mu\nu}q_{i}F^{\mu\nu}+\underset{{\scriptstyle q_{i}}}{{\displaystyle \sum}}\frac{\kappa_{Z}^{q_{i}}}{2\Lambda}}g_{Z}\bar{q_{4}}\sigma_{\mu\nu}q_{i}Z^{\mu\nu}+\underset{{\scriptstyle q_{i}}}{{\displaystyle \sum}}\frac{{\displaystyle \kappa_{g}^{q_{i}}}}{\Lambda}g_{s}\bar{q_{4}}\sigma_{\mu\nu}T^{a}q_{i}G_{a}^{\mu\nu}+H.c.\end{equation}

\hfill{}\hfill{}

where $F^{\mu\nu}$, $Z^{\mu\nu}$ and $G^{\mu\nu}$ are the field
strength tensors of the gauge bosons, $\sigma_{\mu\nu}$ is the anti-symmetric
tensor, $T^{a}$ are Gell-Mann matrices, $e_{q}$ is electric charge
of quark, $g_{e},$ $g_{Z}$ and $g_{s}$ are electromagnetic, neutral
weak and strong coupling constants, respectively. $g_{Z}=g_{e}/cos\theta_{W}sin\theta_{W}$
where $\theta_{W}$ is the Weinberg angle. $\kappa_{\gamma}$, $\kappa_{Z}$
and $\kappa_{g}$ are the strength of anomalous couplings with photon,
$Z$ boson and gluon, respectively. $\Lambda$ is the cutoff scale
for new physics. This type of gauge and Lorentz invariant effective
Lagrangian have been proposed in the framework of composite models
for interactions of excited fermions with ordinary fermions and gauge
bosons \cite{Cabibbo,Hagiwara} . For numerical calculations we implement
the Lagrangian (1), as well as fourth family SM Lagrangian into the
CalcHEP package \cite{Pukhov}. 

The partial decay widths of $u_{4}$ for SM ($u_{4}\rightarrow W^{+}q$
where $q=d$, $s$, $b$) and anomalous ($u_{4}\rightarrow\gamma q$,
$u_{4}\rightarrow Zq$, $u_{4}\rightarrow gq$ where $q=$$u$, $c$,
$t$) modes are given below:

\begin{equation}
\Gamma(u_{4}\rightarrow W^{+}q)=\frac{|V_{u_{4}q}|^{2}\alpha_{e}m_{u_{4}}^{3}}{16m_{{\scriptscriptstyle W}}^{2}sin^{2}\theta_{{\scriptscriptstyle W}}}\varsigma_{{\scriptscriptstyle W}}\sqrt{\varsigma_{{\scriptscriptstyle 0}}}\end{equation}

\medskip{}

where $\varsigma_{{\scriptstyle {\scriptscriptstyle W}}}=(1+x_{q}^{4}+x_{q}^{2}x_{{\scriptscriptstyle W}}^{2}-2x_{q}^{2}-2x_{{\scriptscriptstyle W}}^{4}+x_{{\scriptscriptstyle W}}^{2})$,
$\varsigma_{0}=(1+x_{W}^{4}+x_{q}^{4}-2x_{W}^{2}-2x_{q}^{2}-2x_{W}^{2}x_{q}^{2})$,
$x_{q}=(m_{q}/m_{u_{4}})$ and $x_{{\scriptscriptstyle W}}=(m_{{\scriptscriptstyle W}}/m_{u_{4}})$,

\begin{equation}
\Gamma(u_{4}\rightarrow Zq)=\frac{\alpha_{e}m_{u_{4}}^{3}}{16cos^{2}\theta_{{\scriptscriptstyle W}}sin^{2}\theta_{{\scriptscriptstyle W}}}(\frac{\kappa_{{\scriptscriptstyle Z}}^{q}}{\Lambda})^{2}\varsigma_{{\scriptscriptstyle Z}}\sqrt{\varsigma_{{\scriptscriptstyle 1}}}\end{equation}

\medskip{}

where $\varsigma_{Z}=(2-x_{Z}^{4}-x_{Z}^{2}-4x_{q}^{2}-x_{q}^{2}x_{Z}^{2}-6x_{q}x_{Z}^{2}+2x_{q}^{4})$,
$\varsigma_{1}=(1+x_{Z}^{4}+x_{q}^{2}-2x_{Z}^{2}-2x_{q}^{2}-2x_{Z}^{2}x_{q}^{2})$
and $x_{Z}=(m_{{\scriptscriptstyle Z}}/m_{u_{4}})$,

\begin{equation}
\Gamma(u_{4}\rightarrow gq)=\frac{2\alpha_{s}m_{u_{4}}^{3}}{3}(\frac{\kappa_{g}^{q}}{\Lambda})^{2}\varsigma_{{\scriptscriptstyle 2}}\end{equation}

\medskip{}

where $\varsigma_{2}=(1-3x_{q}^{2}+3x_{q}^{4}-x_{q}^{6})$,

\begin{equation}
\Gamma(u_{4}\rightarrow\gamma q)=\frac{\alpha_{e}m_{u_{4}}^{3}Q_{q}^{2}}{2}(\frac{\kappa_{\gamma}^{q}}{\Lambda})^{2}\varsigma_{{\scriptscriptstyle 2}}.\end{equation}

\medskip{}

The partial decay widths of $d_{4}$ for SM ($d_{4}\rightarrow W^{-}q$
where $q=u$, $c$, $t$) and anomalous ($d_{4}\rightarrow\gamma q$,
$d_{4}\rightarrow Zq$, $d_{4}\rightarrow gq$ where $q=d$, $s$,
$b$) modes are given below:

\begin{equation}
\Gamma(d_{4}\rightarrow W^{-}q)=\frac{|V_{qd_{4}}|^{2}\alpha_{e}m_{d_{4}}^{3}}{16M_{{\scriptscriptstyle W}}^{2}sin^{2}\theta_{{\scriptscriptstyle W}}}\chi_{{\scriptscriptstyle W}}\sqrt{\chi_{{\scriptscriptstyle 0}}}\end{equation}

\medskip{}

where $\chi_{{\scriptstyle {\scriptscriptstyle W}}}=(1+y_{q}^{4}+y_{q}^{2}y_{{\scriptscriptstyle W}}^{2}-2y_{q}^{2}-2y_{{\scriptscriptstyle W}}^{4}+y_{{\scriptscriptstyle W}}^{2})$,
$\chi_{0}=(1+y_{W}^{4}+y_{q}^{4}-2y_{W}^{2}-2y_{q}^{2}-2y_{W}^{2}y_{q}^{2})$,
$y_{q}=(m_{q}/m_{d_{4}})$ and $y_{{\scriptscriptstyle W}}=(m_{{\scriptscriptstyle W}}/m_{d_{4}})$,

\begin{equation}
\Gamma(d_{4}\rightarrow Zq)=\frac{\alpha_{e}m_{d_{4}}^{3}}{16cos^{2}\theta_{{\scriptscriptstyle W}}sin^{2}\theta_{{\scriptscriptstyle W}}}(\frac{\kappa_{{\scriptscriptstyle Z}}^{q}}{\Lambda})^{2}\chi_{{\scriptscriptstyle Z}}\sqrt{\chi_{{\scriptscriptstyle 1}}}\end{equation}

\medskip{}

where $\chi_{Z}=(2-y_{Z}^{4}-y_{Z}^{2}-4y_{q}^{2}-y_{q}^{2}y_{Z}^{2}-6y_{q}y_{Z}^{2}+2y_{q}^{4})$,
$\chi_{1}=(1+y_{Z}^{4}+y_{q}^{2}-2y_{Z}^{2}-2y_{q}^{2}-2y_{Z}^{2}y_{q}^{2})$
and $y_{Z}=(m_{{\scriptscriptstyle Z}}/m_{d_{4}})$,

\begin{equation}
\Gamma(d_{4}\rightarrow gq)=\frac{2\alpha_{s}m_{d_{4}}^{3}}{3}(\frac{\kappa_{g}^{q}}{\Lambda})^{2}\chi_{{\scriptscriptstyle 2}}\end{equation}

\medskip{}

where $\chi_{2}=(1-3y_{q}^{2}+3y_{q}^{4}-y_{q}^{6})$,

\begin{equation}
\Gamma(d_{4}\rightarrow\gamma q)=\frac{\alpha_{e}m_{d_{4}}^{3}Q_{q}^{2}}{2}(\frac{\kappa_{\gamma}^{q}}{\Lambda})^{2}\chi_{{\scriptscriptstyle 2}}.\end{equation}

\medskip{}

One can wonder what is the criteria for the dominance of anomalous
decay modes over SM ones. It is seen from Eq. (6)-(9) that the anomalous
decay modes of the fourth SM family quarks are dominant, i.e. $\Gamma(d_{4}\rightarrow gq)+\Gamma(d_{4}\rightarrow Zq)+\Gamma(d_{4}\rightarrow\gamma q)>\Gamma(d_{4}\rightarrow W^{-}q)$,
if the relation $(\kappa/\Lambda)\gtrsim1.2(V_{ud_{4}}^{2}+V_{cd_{4}}^{2}+V_{td_{4}}^{2})^{1/2}$
TeV$^{-1}$ is satisfied (hereafter $\kappa_{Z}^{q}=\kappa_{g}^{q}=\kappa_{\gamma}^{q}=\kappa$
is assumed). The experimental upper bounds for the fourth family quark
CKM matrix elements are $|V_{u_{4}d}|\leq0.063$, $|V_{u_{4}s}|\leq0.46$,
$|V_{u_{4}b}|\leq0.47$, $|V_{ud_{4}}|\leq0.044$, $|V_{cd_{4}}|\leq0.46$,
$|V_{td_{4}}|\leq0.47$ \cite{Ozcan2008-1}. On the other hand, the
predicted values of these matrix elements are expected to be rather
small in the framework of flavor democracy hypothesis. For example,
the mass matrix parametrization proposed in \cite{Ciftci}, which
gives correct predictions for CKM and MNS mixing matrix elements through
use of SM fermion mass values as input, predicts $|V_{u_{4}d}|=0.0005$,
$|V_{u_{4}s}|=0.0011$, $|V_{u_{4}b}|=0.0014$, $|V_{ud_{4}}|=0.0002$,
$|V_{cd_{4}}|=0.0012$, $|V_{td_{4}}|=0.0014$. In this case, the
anomalous decay modes are dominant, if $(\kappa/\Lambda)>0.0022$
TeV$^{-1}$. The latter correspondens to upper limit $500$ TeV for
new physics scale $\Lambda$, assuming $\kappa=O(1)$.

In Figs 2-5, we plotted branching ratios of $u_{4}$ quark as a function
of $V_{u_{4}b}$ for different values of $\kappa/\Lambda$. Branching
ratios of $u_{4}$ quarks as a function of $\kappa/\Lambda$ for different
values of $V_{u_{4}b}$ are shown Figs. 6-8. It is seen that the assumption
of the dominance of anomalous decay modes is quite realistic, especially
for small CKM mixing parameters. Total decay widths of $u_{4}$ and
$d_{4}$ quarks depending on their masses were plotted in Fig 9 and
10, respectively. 

\begin{figure}
\includegraphics[scale=0.7]{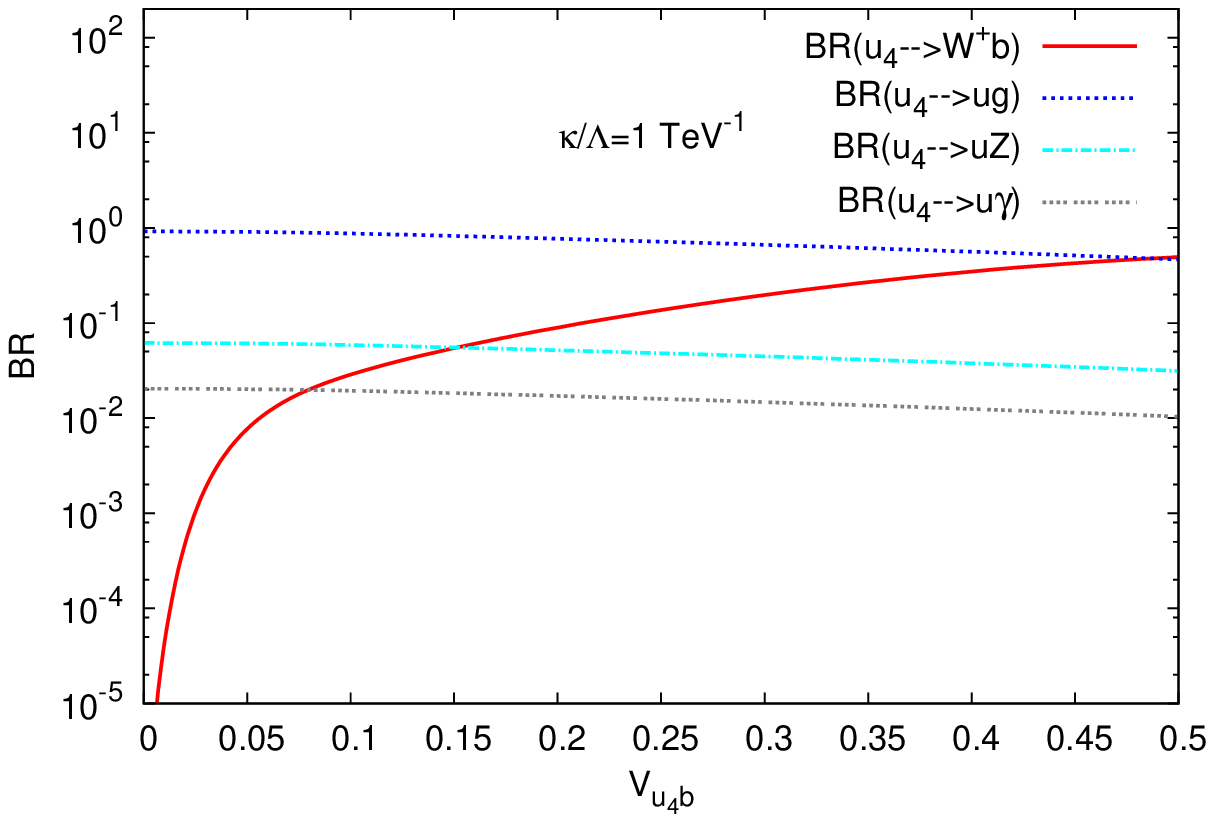}

\caption{Branching ratio of $u_{4}$ versus $V_{u_{4}b}$ graphic for $(\kappa/\Lambda)=1TeV^{-1}$.}

\end{figure}

\begin{figure}
\includegraphics[scale=0.7]{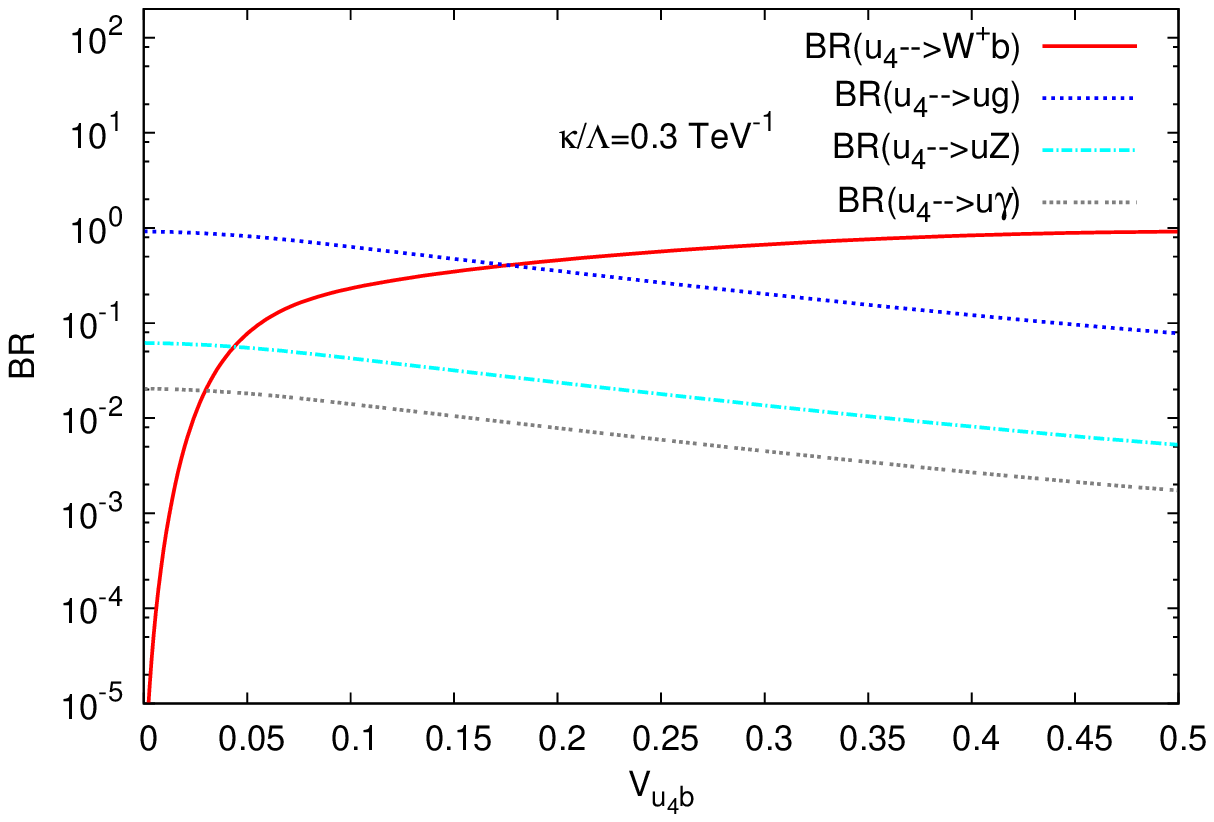}

\caption{Branching ratio of $u_{4}$ versus $V_{u_{4}b}$ graphic for $(\kappa/\Lambda)=0.3TeV^{-1}$}

\end{figure}

\begin{figure}
\includegraphics[scale=0.7]{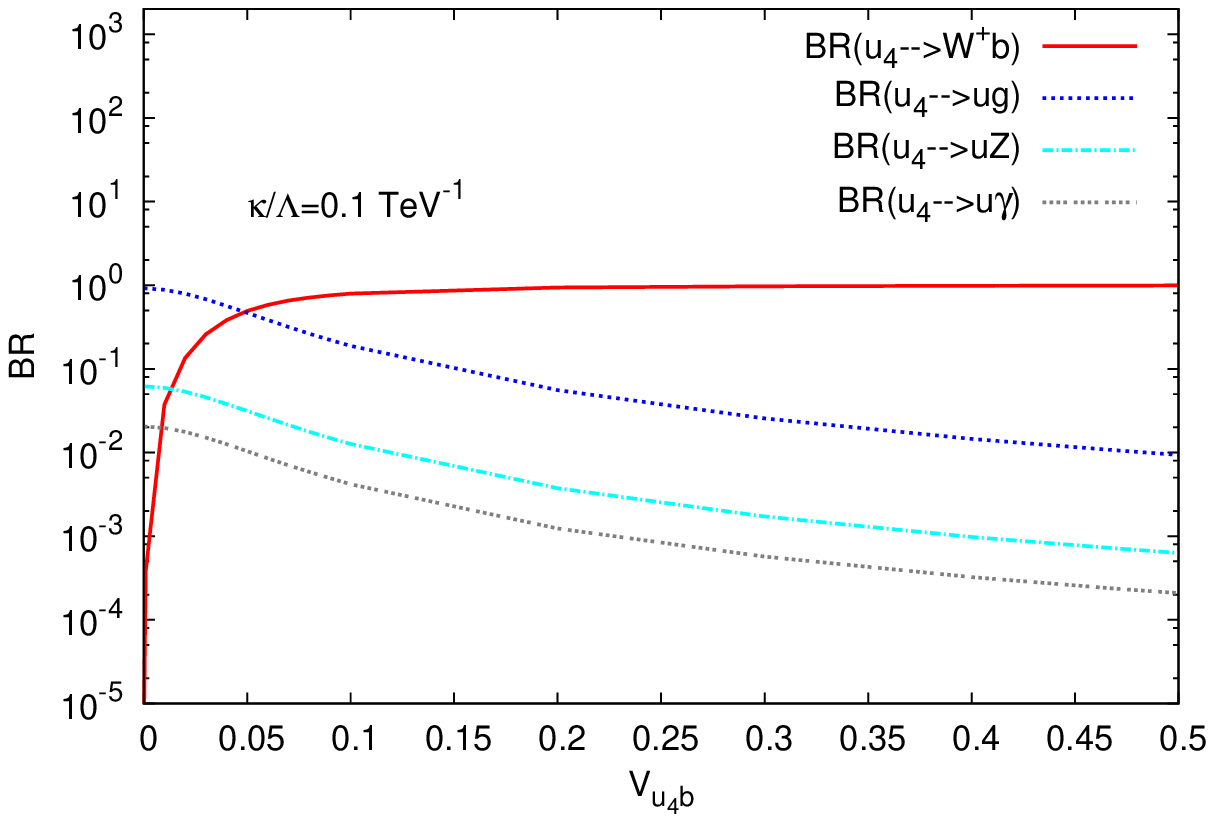}

\caption{Branching ratio of $u_{4}$ versus $V_{u_{4}b}$ graphic for $(\kappa/\Lambda)=0.1TeV^{-1}$}

\end{figure}

\begin{figure}
\includegraphics[scale=0.7]{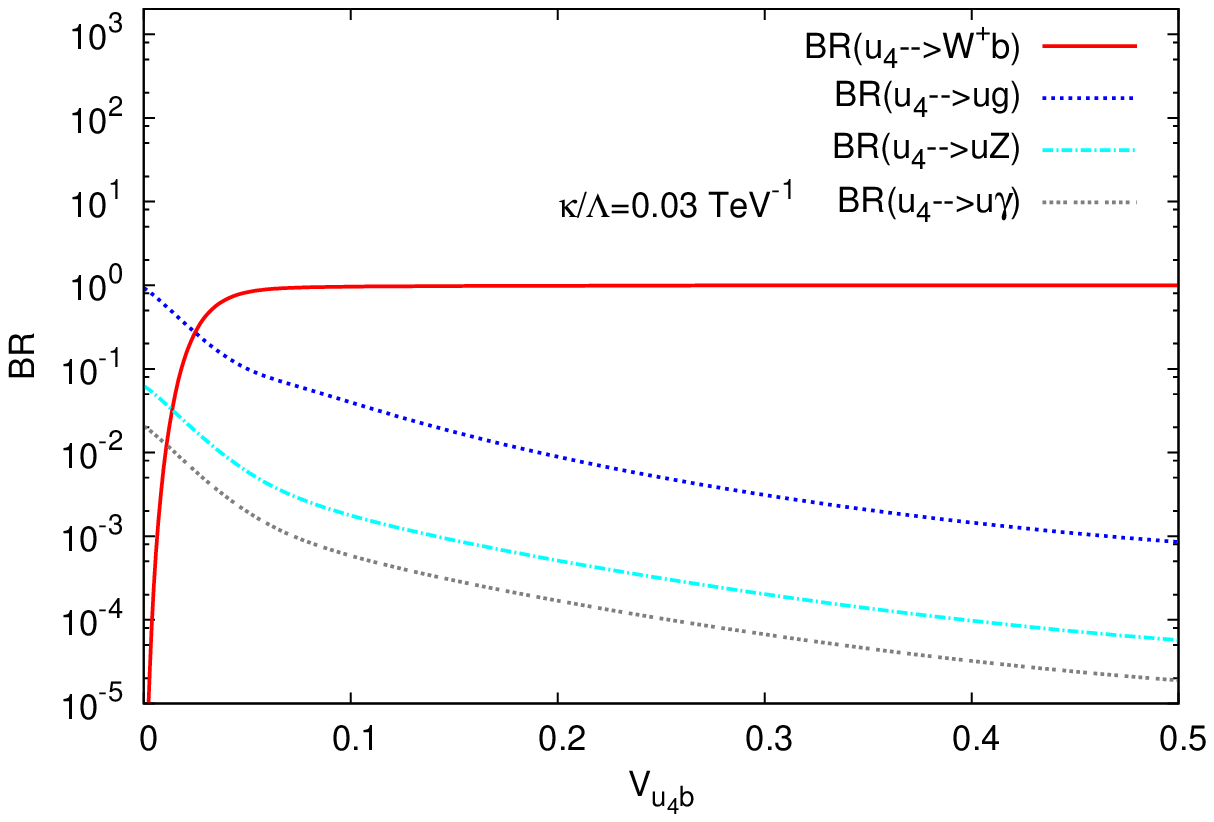}

\caption{Branching ratio of $u_{4}$ versus $V_{u_{4}b}$ graphic for $(\kappa/\Lambda)=0.03TeV^{-1}$}

\end{figure}

\begin{figure}
\includegraphics[scale=0.7]{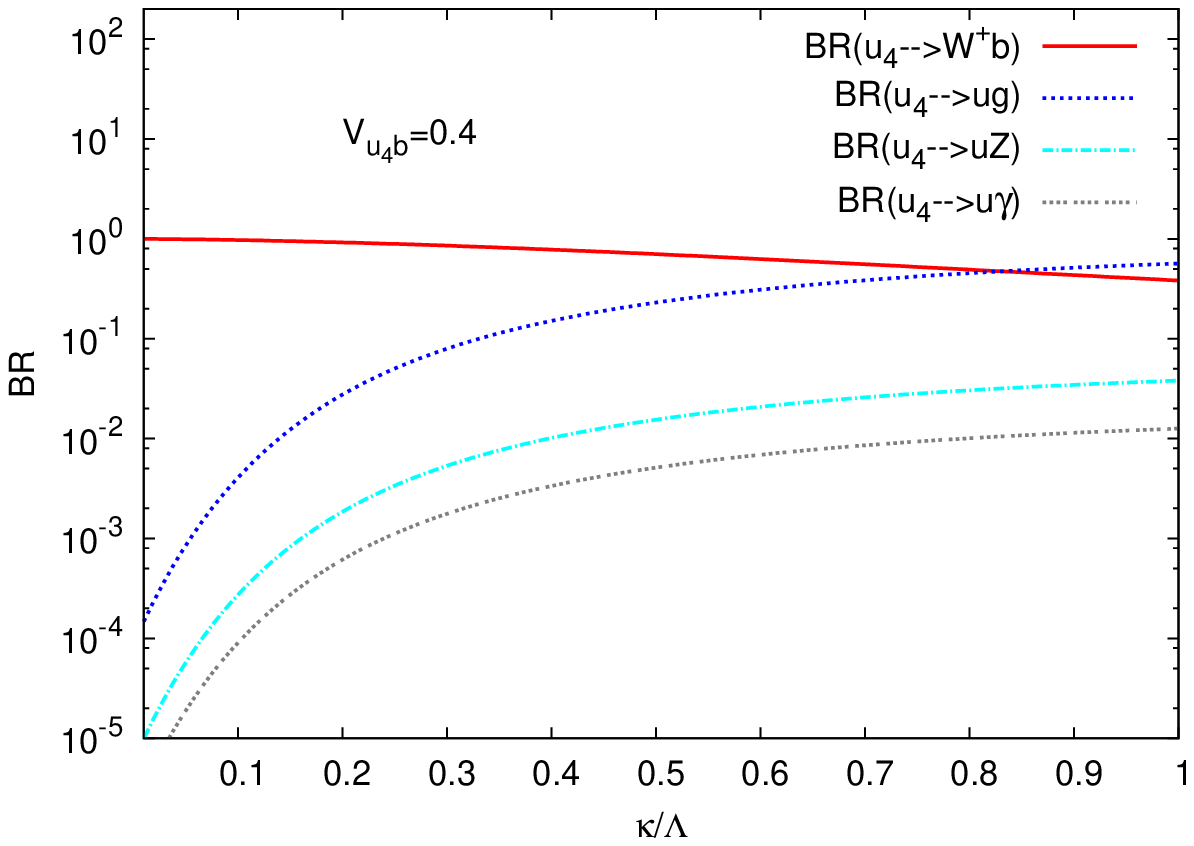}

\caption{Branching ratio of $u_{4}$ versus ($\kappa/\Lambda$) graphic for
$V_{u_{4}b}=0.4$}

\end{figure}

\begin{figure}
\includegraphics[scale=0.7]{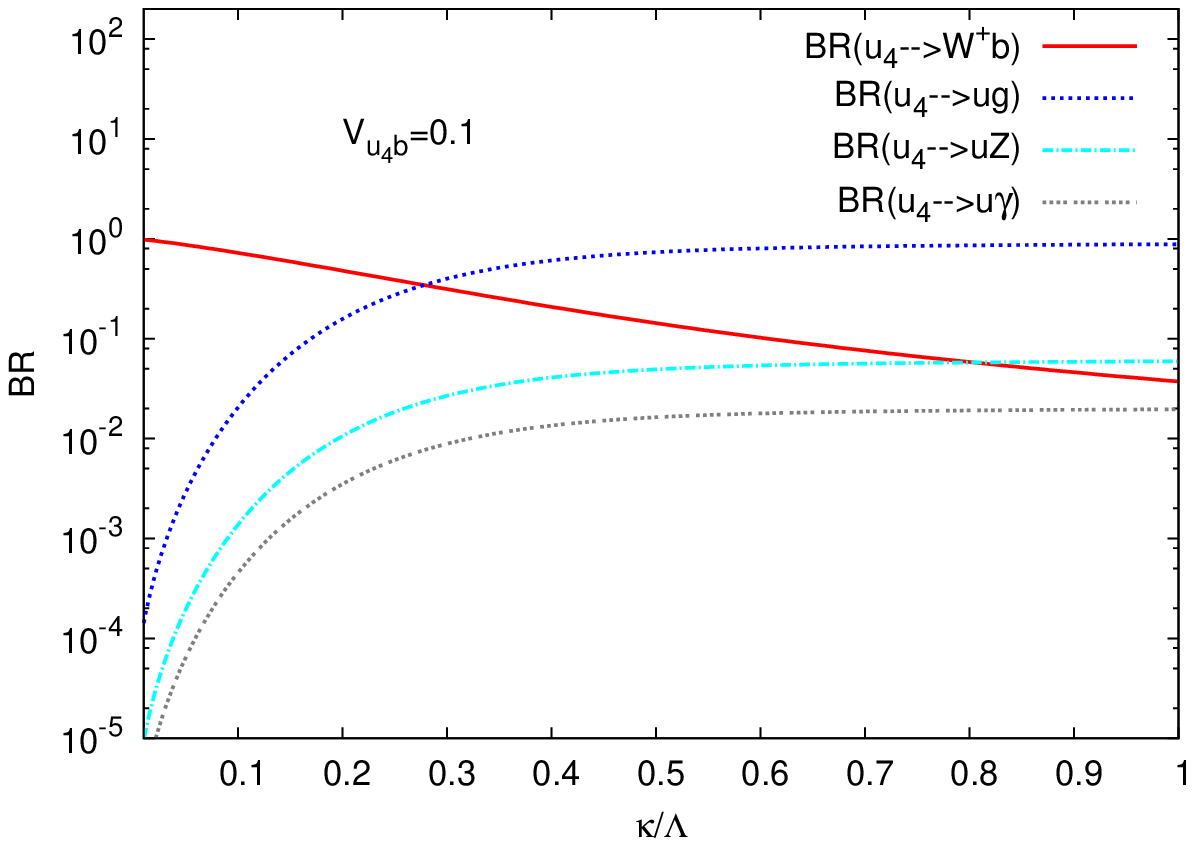}

\caption{Branching ratio of $u_{4}$ versus ($\kappa/\Lambda$) graphic for
$V_{u_{4}b}=0.1$}

\end{figure}

\begin{figure}
\includegraphics[scale=0.7]{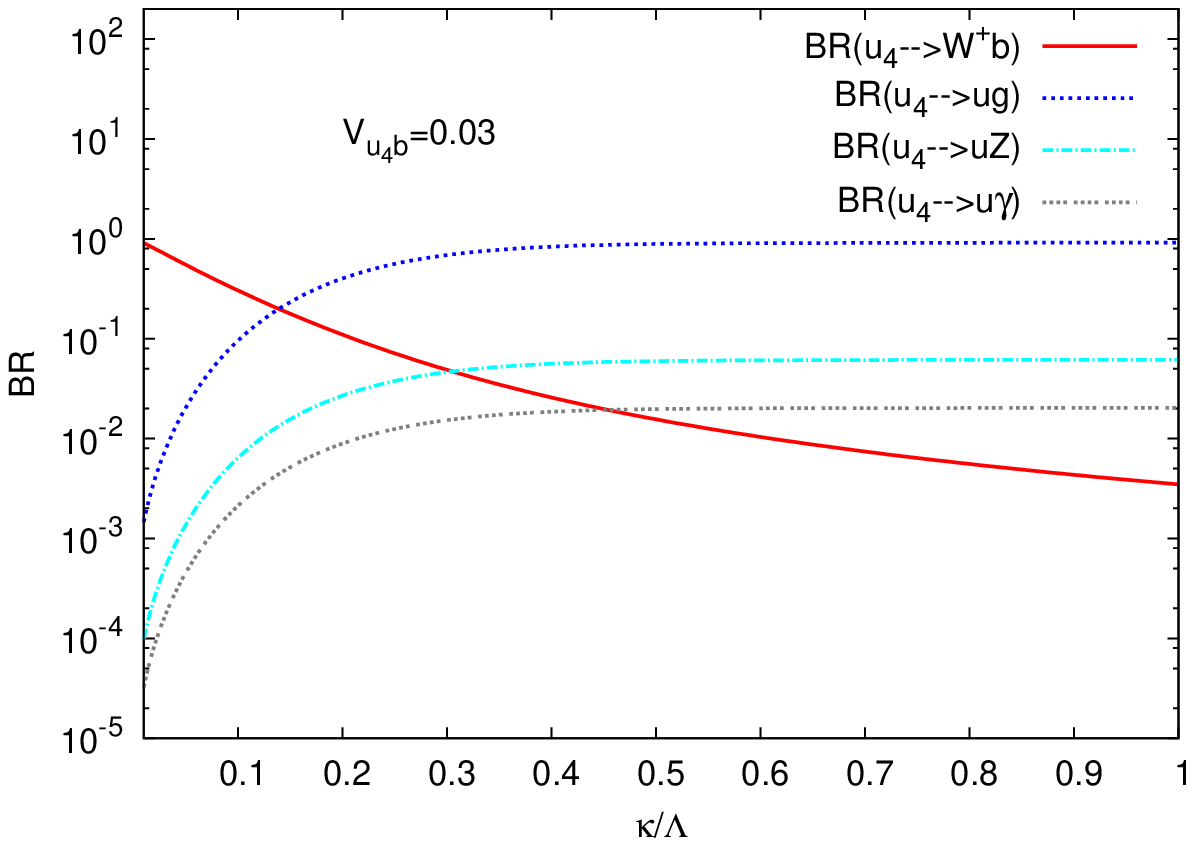}

\caption{Branching ratio of $u_{4}$ versus ($\kappa/\Lambda$) graphic for
$V_{u_{4}b}=0.03$}

\end{figure}

\begin{figure}
\includegraphics[scale=0.7]{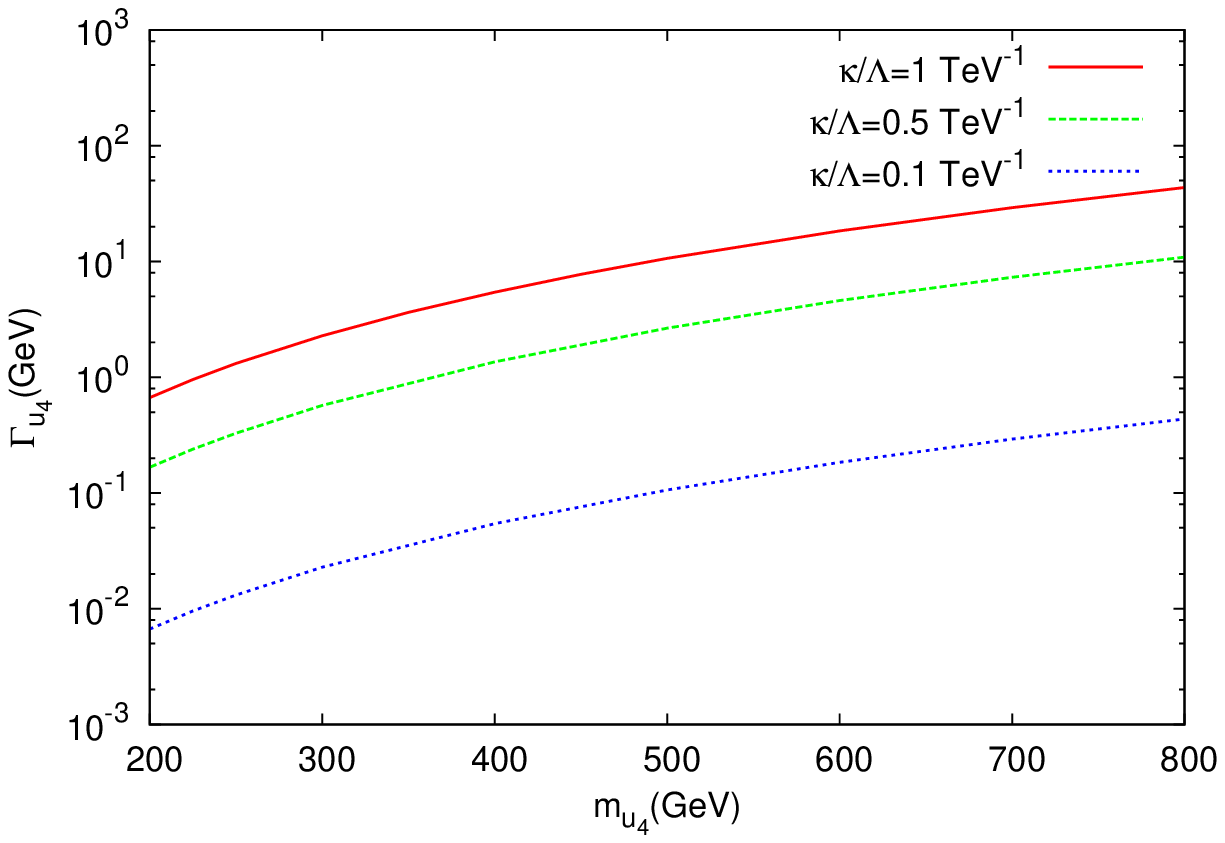}

\caption{Anomalous decay width of $u_{4}$.}

\end{figure}

\begin{figure}
\includegraphics[scale=0.7]{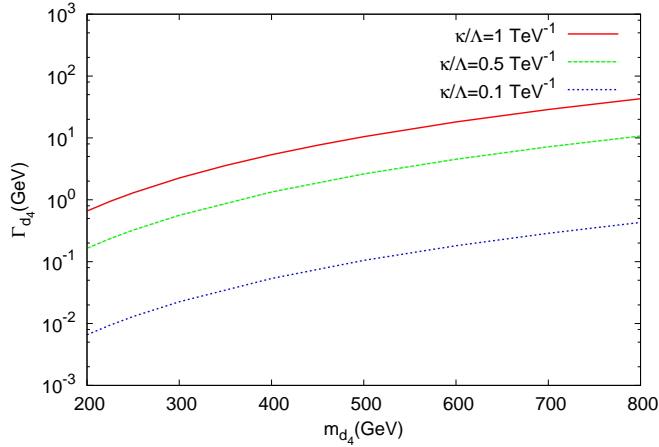}

\caption{Anomalous decay width of $d_{4}$.}

\end{figure}

\section{PAIR $d_{4}$ PRODUCTION AT TEVATRON AND LHC WITH SUBSEQUENT ANOMALOUS
DECAYS}

In this section, we study pair production of $d_{4}$ quarks at the
Tevatron and LHC. For the numerical calculations we implement anomalous
interaction lagrangian of the fourth family quarks into CalcHEP package
program \cite{Pukhov} and we used CTEQ6L \cite{CTEQ6L_Pumplin_Stump}
parton distribution functions with factorization scale $Q^{2}=m_{d_{4}}^{2}$.
The pair production cross sections of $d_{4}\overline{d}_{4}$ at
the Tevatron and LHC are plotted in Fig. 11. It is seen that i.e.
for $m_{d_{4}}=300$ GeV pair production cross section at LHC with
$\sqrt{s}=7$ TeV is $20$ times larger than at Tevatron. This ratio
can be used to compaire Tevatron and LHC capacities. Namely, for $m_{d_{4}}=300$
GeV LHC need twenty times less luminosity than the Tevatron. 

\begin{figure}
\includegraphics[scale=0.7]{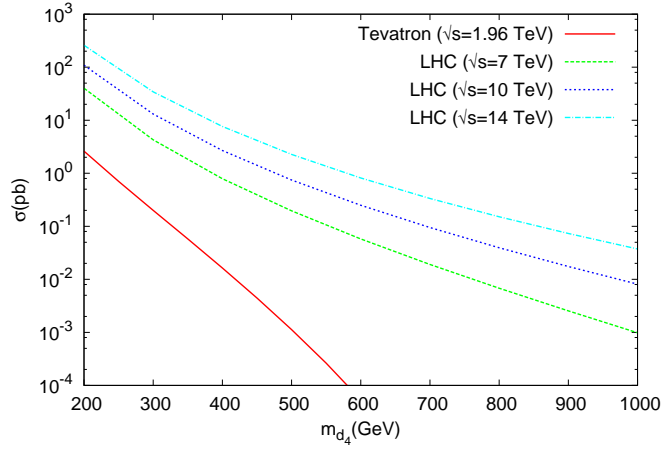}

\caption{The pair production cross section of $d_{4}\overline{d}_{4}$ at the
Tevatron and LHC}

\end{figure}

Pair production of fourth SM family quarks at hadron colliders have
been analysed in a number of papers (see corresponding rows in the
Table VI ) assuming SM decays. For this reason below we consider the
process $p\overline{p}(p)\rightarrow d_{4}\overline{d}_{4}X\rightarrow gd\gamma\overline{d}X$
in order to compare the Tevatron and LHC search potential in the case
where anomalous decays are dominant. This process will be seen in
dedector as $\gamma+3j$ events, for background calculations we use
MADGRAPH package \cite{Alwall}.

\subsection{Signal and Background Analysis at the Tevatron}

Normalized transverse momentum ($p_{T}$) and pseudo-rapidity ($\eta$)
distributions of final state partons (quarks, photon and gluon) for
signal and background processes are shown in Fig. 12 and Fig. 13,
respectively. It is seen that $p_{T}>50$ GeV cut essentially reduces
background, whereas signal is almost unaffected. In addition to $p_{T}>50$
GeV, we have used CDF cut value $|\eta|<2$ for pseudo-rapidity, as
well as invariant mass within $\pm20$ GeV around $d_{4}$ mass. In
table VIII we present the values of the signal and background cross-sections
for different cuts.

\begin{figure}
\includegraphics[scale=0.7]{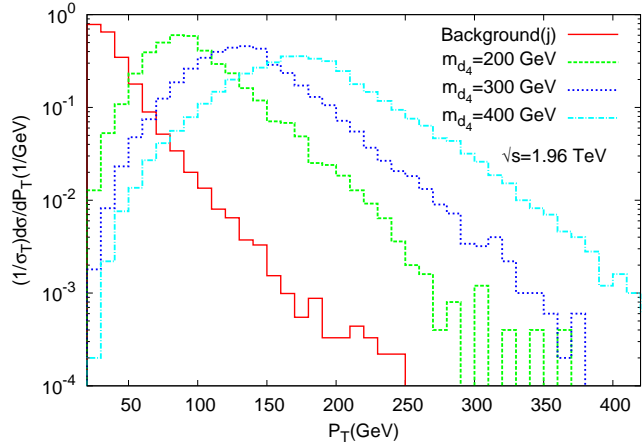}

\caption{Normalised $p_{T}$ distributions of partons for signal and background
for pair $d_{4}$ production at the Tevatron. }

\end{figure}

\begin{figure}
\includegraphics[scale=0.7]{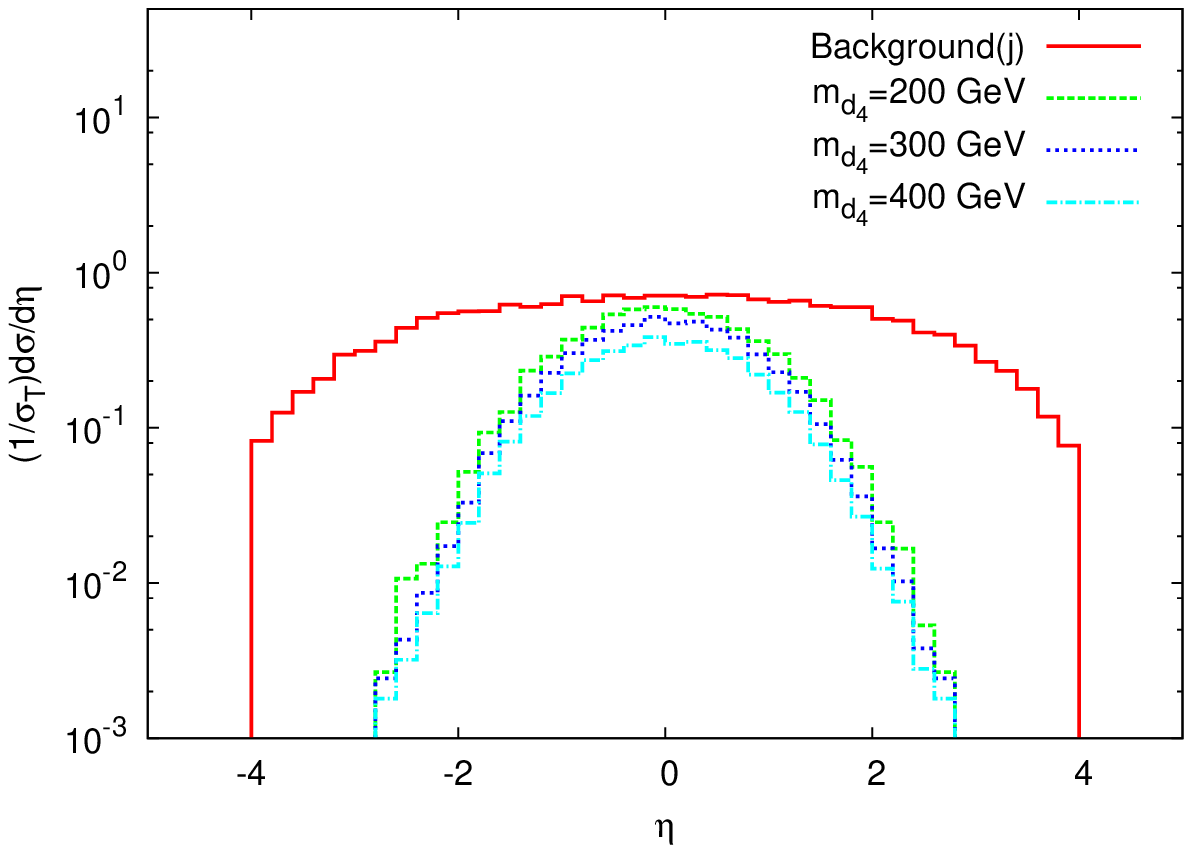}

\caption{Normalized $\eta$ distributions of partons of signal and background
for pair $d_{4}$ production at the Tevatron. }

\end{figure}

\begin{table}
\begin{tabular}{|c|c|c|c|c|c|c|}
\hline 
$M_{d_{4}}$  & \multicolumn{2}{c||}{$200$ GeV} & \multicolumn{2}{c||}{$300$ GeV} & \multicolumn{2}{c||}{$400$ GeV}\tabularnewline
\hline
\hline 
cuts  & $\sigma_{S}$, fb  & $\sigma_{B}$, fb  & $\sigma_{S}$, fb  & $\sigma_{B}$, fb  & $\sigma_{S}$, fb  & $\sigma_{B}$, fb\tabularnewline
\hline 
$p_{T}>20$GeV  & $39.2$  & $5.4\times10^{5}$  & $2.92$  & $5.4\times10^{5}$  & $0.23$  & $5.4\times10^{5}$\tabularnewline
\hline 
$p_{T}>50$GeV  & $24.5$  & $2.7\times10^{3}$  & $2.40$  & $2.7\times10^{3}$  & $0.21$  & $2.7\times10^{3}$\tabularnewline
\hline 
all cuts  & $21.8$  & $3.63$  & $2.27$  & $0.091$  & $0.20$  & $0.006$\tabularnewline
\hline
\end{tabular}

\caption{Signal and background cross sections values for various cuts at Tevatron
\cite{biz}. All cuts include $p_{T}>50$ GeV, $|\eta|<2$, $|M_{inv}(\gamma j)-M_{d_{4}}|<20$
GeV, $|M_{inv}(jj)-M_{d_{4}}|<20$ GeV.}

\end{table}

Statistical significance has been calculated by using following formula
\cite{CMS_Note_significance}:

\begin{equation}
S=\sqrt{2[(s+b)ln(1+\frac{s}{b})-s]}\end{equation}

where $s$ and $b$ represents the numbers of signal and background
events, respectively.

In Fig. 14 we plot the neccesary luminosity for $2\sigma$ exclusion,
$3\sigma$ observation and $5\sigma$ discovery limits depending on
$d_{4}$ mass. Reachable masses for $d_{4}$ quark at different values
of the Tevatron integrated luminosity are presented in Table IX.

\begin{figure}
\includegraphics[scale=0.7]{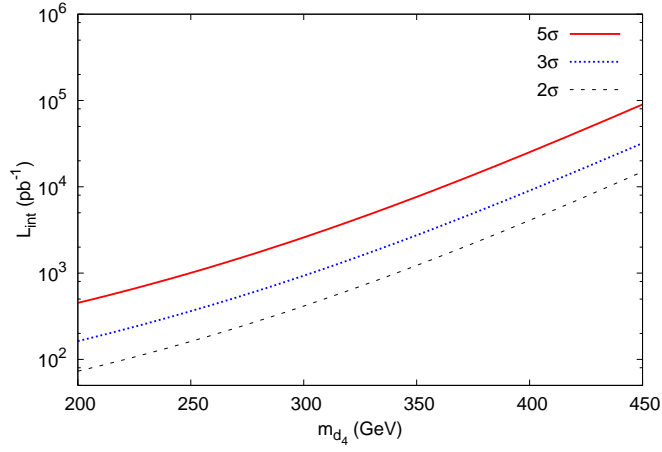}

\caption{The necessary integrated luminosity for exclusion, observation and
discovery of $d_{4}$ quark at the Tevatron \cite{biz}.}

\end{figure}

\begin{table}
\begin{tabular}{|c|c|c|c|}
\hline 
$L_{int}$, $fb^{-1}$ & $5$ & $10$ & $20$\tabularnewline
\hline
\hline 
$2\sigma$ exclusion & $390$ GeV & $430$ GeV & $460$ GeV\tabularnewline
\hline 
$3\sigma$ observation & $370$ GeV & $410$ GeV & $440$ GeV\tabularnewline
\hline 
$5\sigma$ discovery & $340$ GeV & $360$ GeV & $390$ GeV\tabularnewline
\hline
\end{tabular}

\caption{Reachable $m_{d_{4}}$ mass values for discovery, observation, and
exclusion at the Tevatron \cite{biz}.}

\end{table}

\subsection{Signal and Background Analysis at the LHC}

Normalized transverse momentum ($p_{T}$) and pseudo-rapidity ($\eta$)
distributions of final state partons (quarks, photon and gluon) for
signal and background processes are shown in Fig. 15 and Fig. 16,
respectively. It is seen that $p_{T}>50$ GeV cut essentially reduces
background, whereas signal is almost unaffected. In addition to $p_{T}>50$
GeV, we have used ATLAS cut value $|\eta|<2.5$ for pseudo-rapidity,
as well as invariant mass within $\pm20$ GeV around $d_{4}$ mass.
In Table X we present the values of the signal and background cross-sections
for different cuts.

\begin{figure}
\includegraphics[scale=0.7]{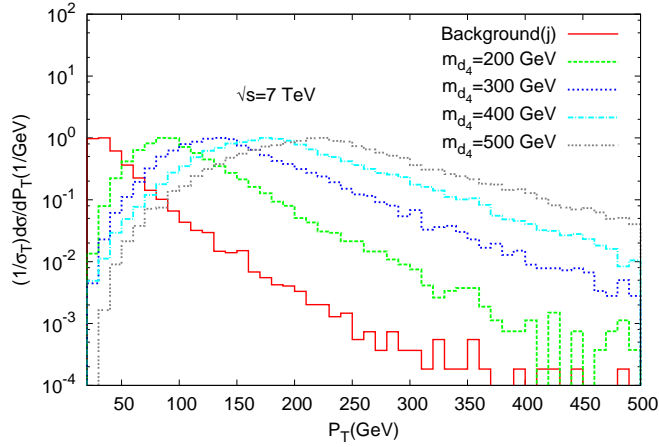}

\caption{Normalised $p_{T}$ distributions of partons for the signal and background
for pair $d_{4}$ production at the LHC.}

\end{figure}

\begin{figure}
\includegraphics[scale=0.7]{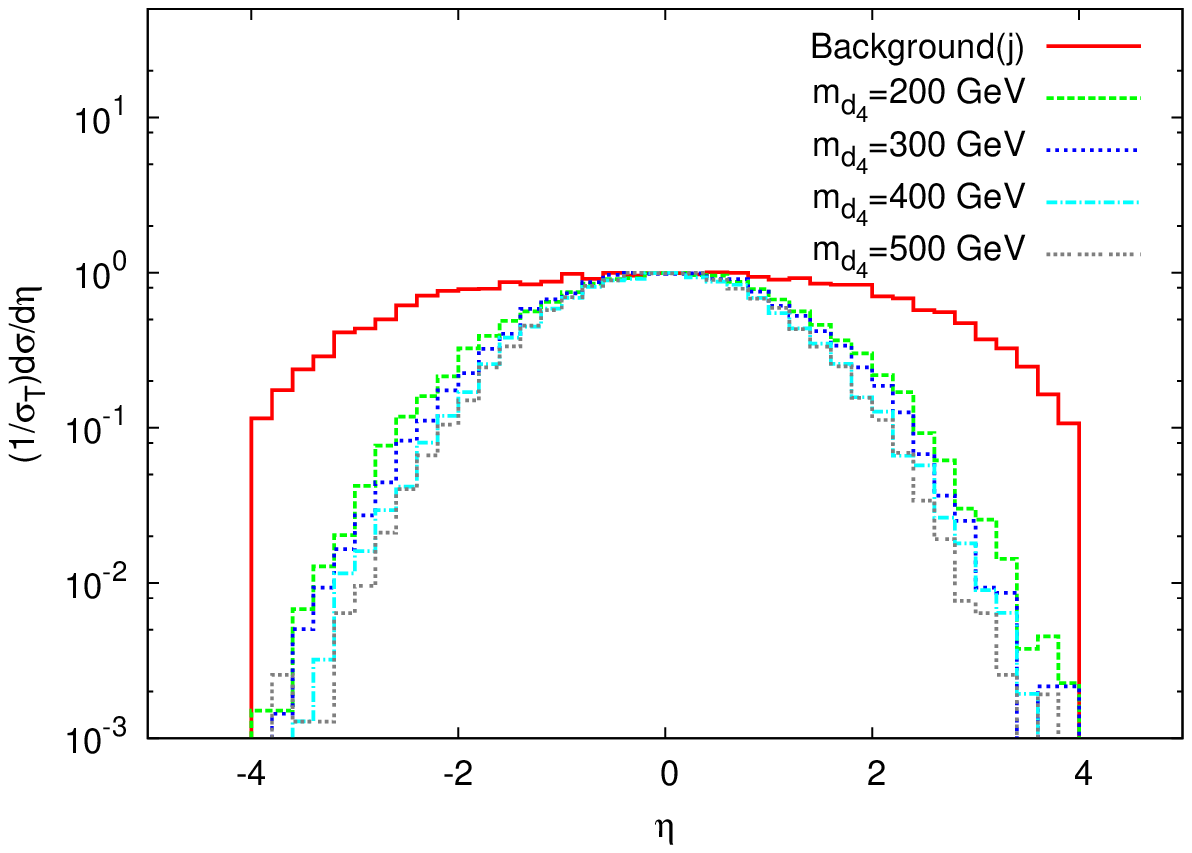}

\caption{Normalised $\eta$ distributions of partons for the signal and background
for pair $d_{4}$ production at the LHC.}

\end{figure}

\begin{table}
\begin{tabular}{|c|c|c|c|c|c|c|l|l|}
\hline 
$M_{d_{4}}$  & \multicolumn{2}{c||}{$200$ GeV} & \multicolumn{2}{c||}{$300$ GeV} & \multicolumn{2}{c||}{$400$ GeV} & \multicolumn{2}{l|}{500 GeV}\tabularnewline
\hline
\hline 
cuts  & $\sigma_{S}$, fb  & $\sigma_{B}$, fb  & $\sigma_{S}$, fb  & $\sigma_{B}$, fb  & $\sigma_{S}$, fb  & $\sigma_{B}$, fb & $\sigma_{S}$, fb  & $\sigma_{B}$, fb\tabularnewline
\hline 
$p_{T}>20$ GeV  & $3.77\times10^{3}$  & $7.44\times10^{6}$  & $394$  & $7.44\times10^{6}$  & $71.1$  & $7.44\times10^{6}$  & $19.2$ & $7.44\times10^{6}$ \tabularnewline
\hline 
$p_{T}>50$ GeV  & $2.14\times10^{3}$  & $1.12\times10^{5}$  & $319$  & $1.12\times10^{5}$  & $63.7$  & $1.12\times10^{5}$  & $17.9$ & $1.12\times10^{5}$ \tabularnewline
\hline 
all cuts  & $315$  & $13.62$  & $46.94$  & $1.03$  & $9.3$  & $0.59$ & $2.4$ & $0.037$\tabularnewline
\hline
\end{tabular}

\caption{Signal and background cross sections values for various cuts at the
LHC. All cuts include $p_{T}>50$ GeV, $|\eta|<2.5$, $|M_{inv}(\gamma j)-M_{d_{4}}|<20$
GeV, $|M_{inv}(jj)-M_{d_{4}}|<20$ GeV.}

\end{table}

In Fig. 17 we plot the neccesary luminosity for $2\sigma$ exclusion,
$3\sigma$ observation and $5\sigma$ discovery limits depending on
$d_{4}$ mass. Reachable masses for $d_{4}$ quark at different values
of the Tevatron integrated luminosity are presented is Table XI.

\begin{figure}
\includegraphics[scale=0.7]{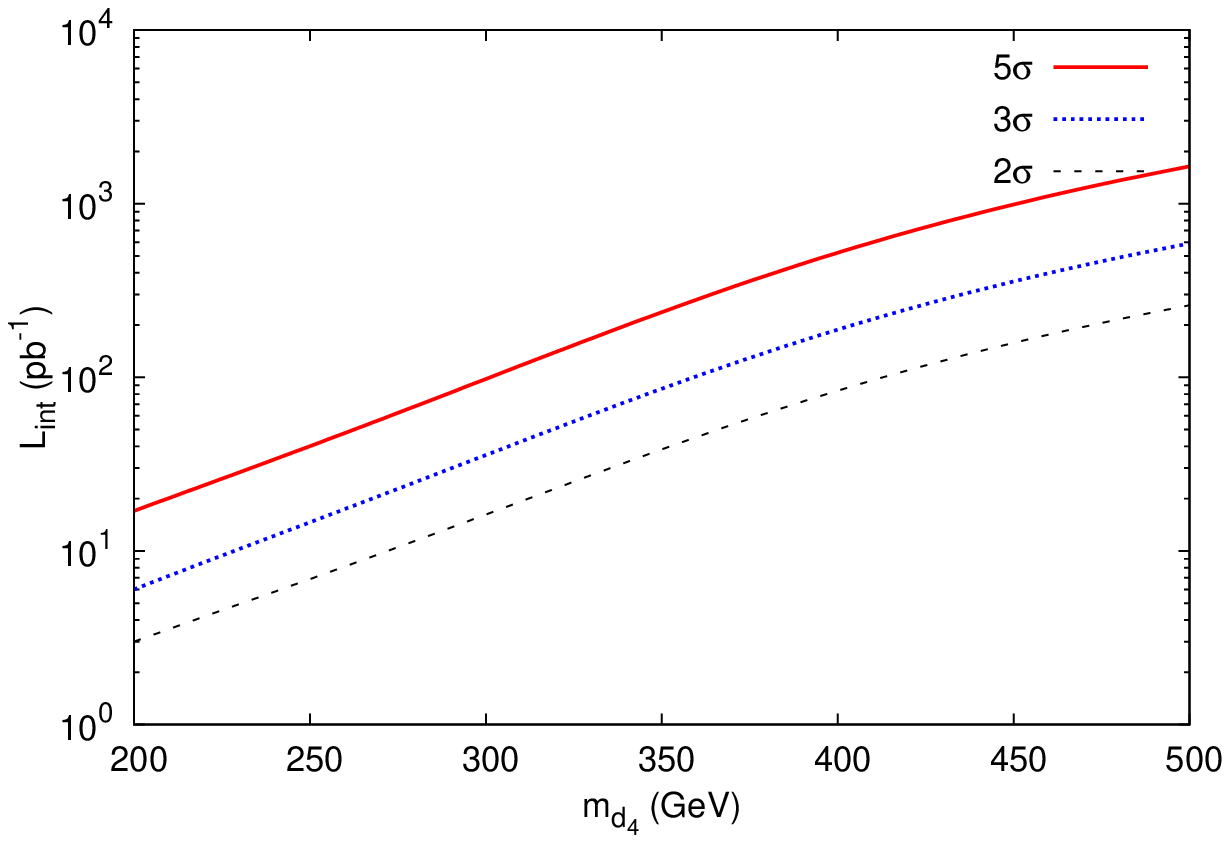}

\caption{The necessary integrated luminosity for exclusion, observation and
discovery of $d_{4}$ quark at the LHC.}

\end{figure}

\begin{table}
\begin{tabular}{|c|c|c|c|}
\hline 
$L_{int}$, $pb^{-1}$ & $100$ & $300$ & $1000$\tabularnewline
\hline
\hline 
$2\sigma$ exclusion & $420$ GeV & $510$ GeV & $640$ GeV\tabularnewline
\hline 
$3\sigma$ observation & $370$ GeV & $430$ GeV & $550$ GeV\tabularnewline
\hline 
$5\sigma$ discovery & $310$ GeV & $360$ GeV & $460$ GeV\tabularnewline
\hline
\end{tabular}

\caption{Reachable $m_{d_{4}}$ mass values for discovery, observation, and
exclusion at the LHC.}

\end{table}

Comparing Tables IX and XI, one can conclude that LHC with $\sqrt{s}=7$
TeV and integrated luminosity $L_{int}=300$ $pb^{-1}$surpasses the
Tevatron with $L_{int}=10fb^{-1}$.

\section{ANOMALOUS RESONANT $u_{4}$ PRODUCTION AT TEVATRON AND LHC WITH SUBSEQUENT
ANOMALOUS DECAY }

Total cross sections for the anomalous resonant production of $u_{4}$
quark at the Tevatron and LHC are shown in Fig. 18. It is seen that
for $m_{d_{4}}=300$ GeV the cross section at the LHC with $\sqrt{s}=7$
TeV is $40$ times larger than the cross section at the Tevatron. 

\begin{figure}
\includegraphics[scale=0.7]{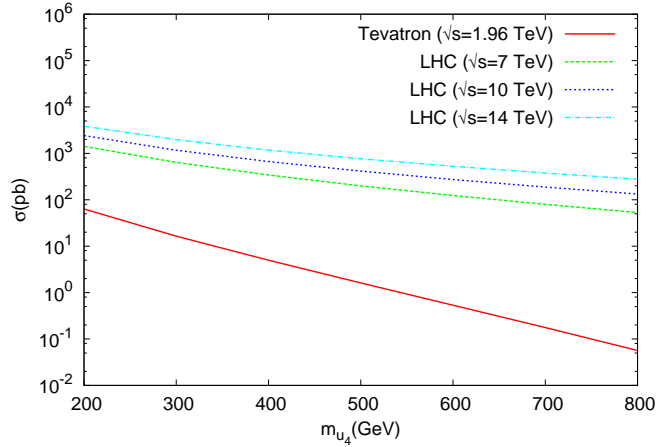}

\caption{The anomalous resonant production cross sections of $u_{4}$ at Tevatron
and LHC.}

\end{figure}

\subsection{Signal and Background Analysis at the Tevatron}

The $p\overline{p}\rightarrow u_{4}X\rightarrow\gamma uX$ process
is considered as a signature of anomalous resonant production of fourth
SM family up type quark. The SM background for this process is $p\overline{p}\rightarrow\gamma jX$,
where $j=u,\overline{u},d,\overline{d},c,\overline{c},s,\overline{s},b,\overline{b},g$.
In order to determine appropriate kinematical cuts, $p_{T}$ and $\lyxmathsym{\textgreek{h}}$
distributions for signal and background processes are given in Fig.
19 and Fig. 20, respectively.

\begin{figure}
\includegraphics[scale=0.7]{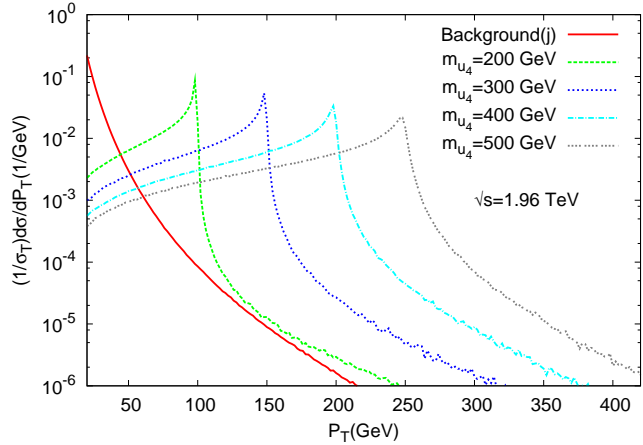}

\caption{Normalised $p_{T}$ distributions of partons for the signal and background
for anomalous resonant $u_{4}$ production at the Tevatron.}

\end{figure}

\begin{figure}
\includegraphics[scale=0.7]{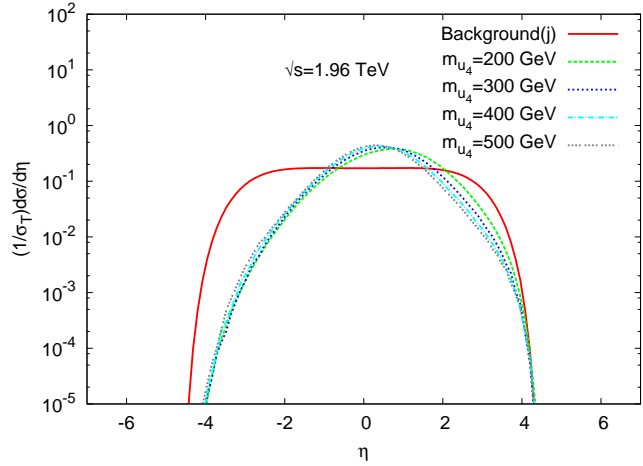}

\caption{Normalised $\eta$ distributions of partons for the signal and background
for anomalous resonant $u_{4}$ production at the Tevatron.}

\end{figure}

In order to extract the $u_{4}$ signal and to suppress the background,
the following cuts are applied: $p_{T}>75$ GeV and $|\eta|<2$ for
all final state partons and photon, as well as invariant mass within
$\pm20$ GeV around the $u_{4}$ mass. For the signal calculations
$\kappa/\Lambda=0.1TeV^{-1}$ have been used. 

In Fig. 21 we plot the neccesary luminosity for $2\sigma$ exclusion,
$3\sigma$ observation and $5\sigma$ discovery limits depending on
$u_{4}$ mass. Reachable masses for $u_{4}$ quark at different values
of the Tevatron integrated luminosity are presented in Table XII.

\begin{figure}
\includegraphics[scale=0.7]{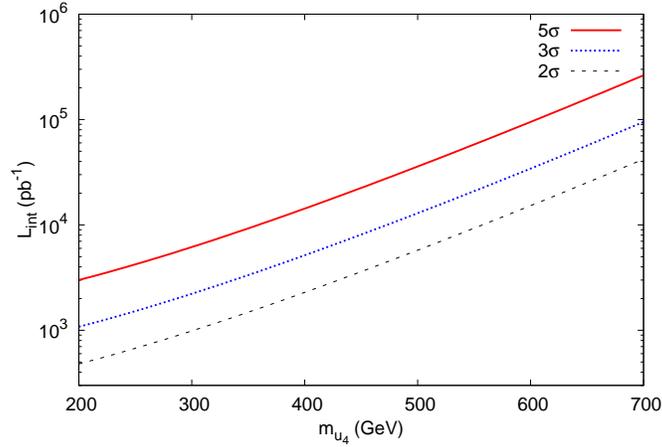}

\caption{The necessary integrated luminosity for exclusion, observation and
discovery of $u_{4}$ quark at the Tevatron}

\end{figure}

\begin{table}
\begin{tabular}{|c|c|c|c|}
\hline 
$L_{int}$, $fb^{-1}$ & $5$ & $10$ & $20$\tabularnewline
\hline
\hline 
$2\sigma$ exclusion & $480$ GeV & $560$ GeV & $630$ GeV\tabularnewline
\hline 
$3\sigma$ observation & $400$ GeV  & $470$ GeV & $540$ GeV\tabularnewline
\hline 
$5\sigma$ discovery & $270$ GeV & $360$ GeV & $440$ GeV\tabularnewline
\hline
\end{tabular}

\caption{Reachable $m_{u_{4}}$ mass values for discovery, observation and
exclusion at the Tevatron}

\end{table}

\subsection{Signal and Background Analysis at the LHC}

In order to determine appropriate kinematical cuts, $p_{T}$ and $\lyxmathsym{\textgreek{h}}$
distributions for signal and background processes are given in Fig.
22 and Fig. 23, respectively.

\begin{figure}
\includegraphics[scale=0.7]{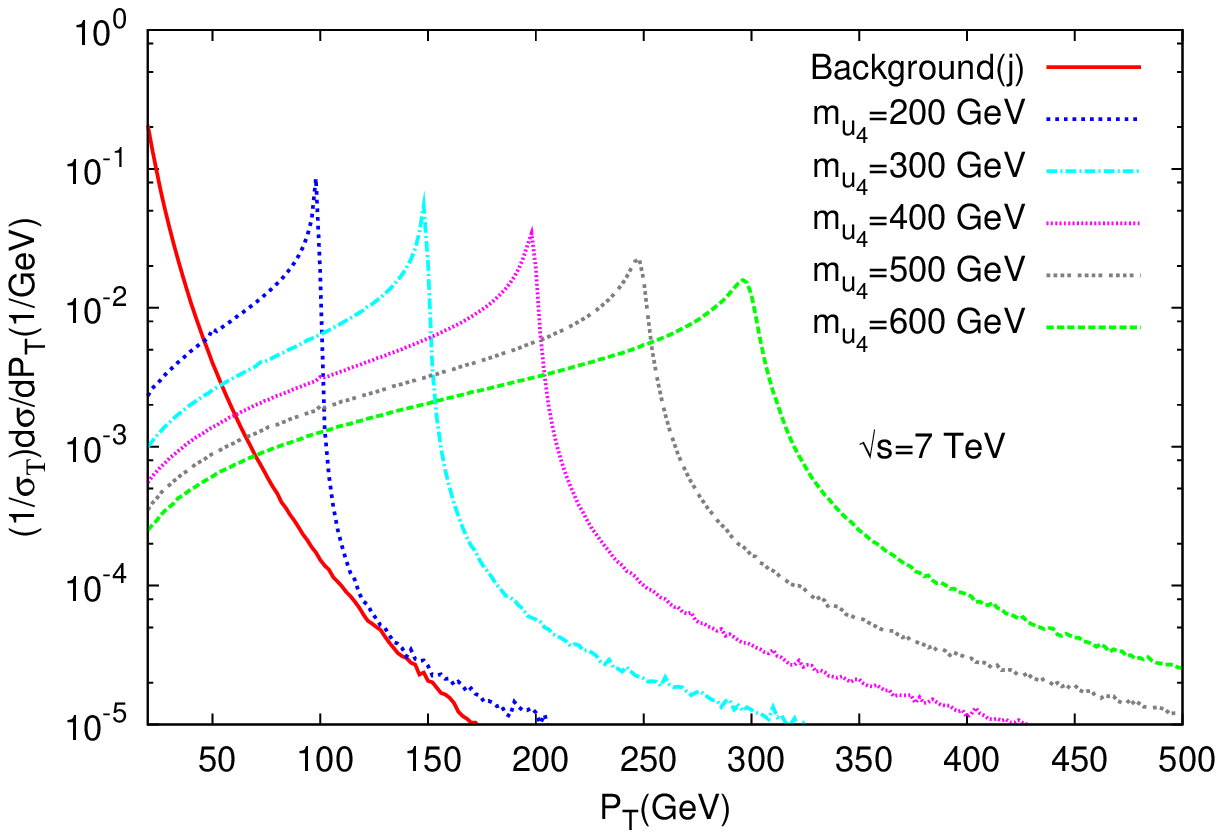}

\caption{Normalised $p_{T}$ distributions of partons for the signal and background
for anomalous resonant $u_{4}$ production at the LHC.}

\end{figure}

\begin{figure}
\includegraphics[scale=0.7]{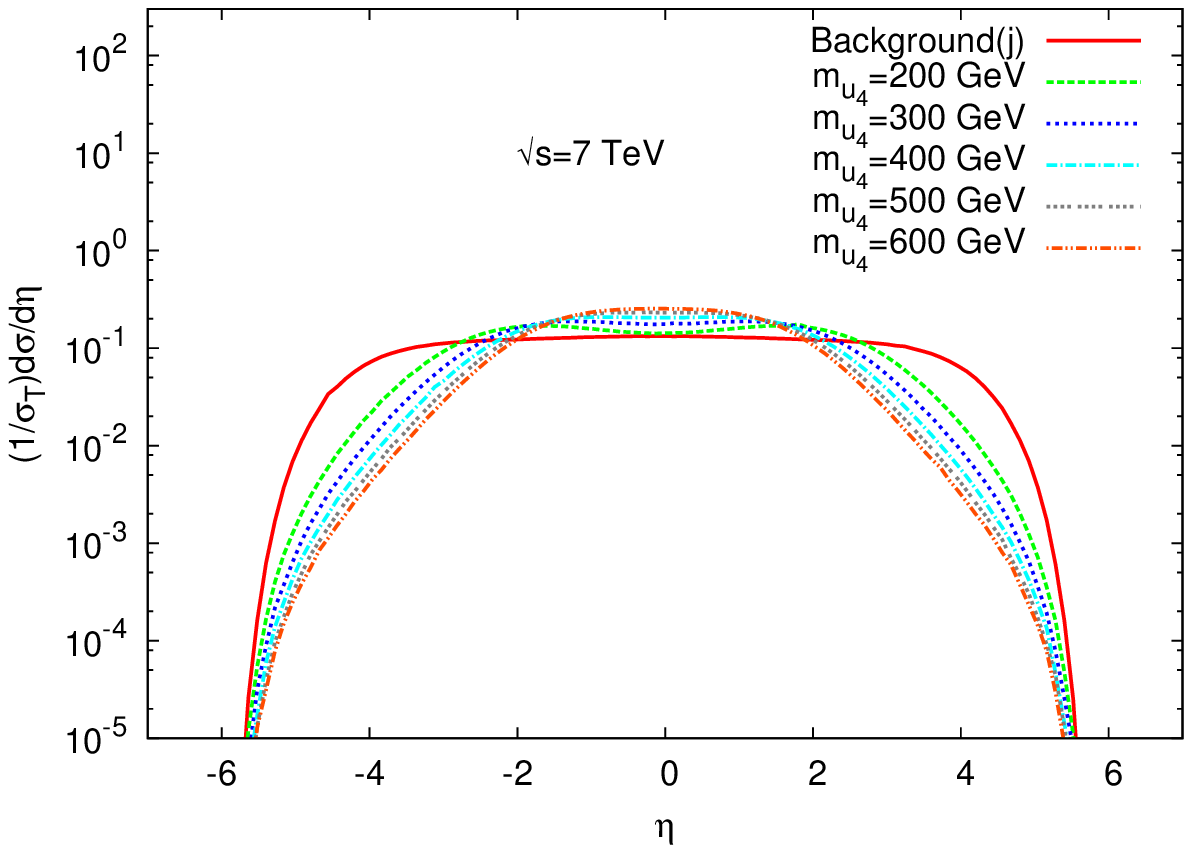}

\caption{Normalised $\eta$ distributions of partons for the signal and background
for anomalous resonant $u_{4}$ production at the LHC.}

\end{figure}

In order to extract the $u_{4}$ signal and to suppress the background,
the following cuts are applied: $p_{T}>75$ GeV and $|\eta|<2.5$
for all final state partons and photon, as well as invariant mass
within $\pm20$ GeV around the $u_{4}$ mass. For the signal calculations
$\kappa/\Lambda=0.1$ $TeV^{-1}$ have been used. 

In Fig. 24 we plot the neccesary luminosity for $2\sigma$ exclusion,
$3\sigma$ observation and $5\sigma$ discovery limits depending on
$u_{4}$ mass. 

Comparing figures 21 and 24, it is obvious that LHC with $\sqrt{s}=7$
TeV and integrated luminosity $L_{int}=100$ $pb^{-1}$surpasses the
Tevatron with $L_{int}=10$ $fb^{-1}$.

\begin{figure}
\includegraphics[scale=0.7]{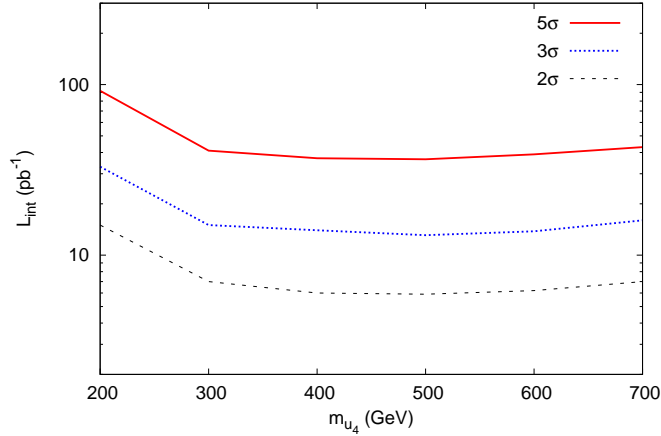}

\caption{The necessary integrated luminosity for exclusion, observation and
discovery of $u_{4}$ quark at the LHC}

\end{figure}

\section{CONCULUSION}

It is seen that there is a though competition between Tevatron and
LHC in a search for Higgs boson and fourth family quarks. We have
shown that in case the anomalous decay modes are dominant:

a) for pair production, LHC with $\sqrt{s}=7$ TeV and $L_{int}=300$
$pb^{-1}$ surpasses Tevatron with $L_{int}=10$ $fb^{-1}$,

b) for anomalous resonant production, LHC with $L_{int}=100$ $pb^{-1}$
cover whole mass range if $\kappa/\Lambda=0.1$ $TeV^{-1}$.

\end{document}